\documentclass[12pt]{iopart}

\usepackage{stmaryrd}

\usepackage{color}
\usepackage{amssymb}
\usepackage{graphicx}
\usepackage{mathrsfs}
\usepackage{CJK}
\usepackage[figuresright]{rotating}
\usepackage{epstopdf}
\usepackage[colorlinks=black,
            linkcolor=black,
            anchorcolor=black,
            citecolor=black]{hyperref}

\usepackage{url}
\usepackage[square, comma, sort&compress, numbers]{natbib}

\newtheorem{asp}{Assumption}[section]
\newtheorem{thm}{Theorem}[section]
\newtheorem{cor}{Corollary}[section]
\newtheorem{lem}{Lemma}[section]
\newtheorem{rmk}{Remark}[section]

\newcommand{\bx}{\mathbf{x}}
\newcommand{\by}{\mathbf{y}}
\newcommand{\be}{\mathbf{e}}
\newcommand{\bu}{\mathbf{u}}
\newcommand{\tu}{\tau}
\newcommand{\ph}{\phi}
\newcommand{\p}{\psi}

\begin{document}

\title[Inferring topologies via driving-based generalized  synchronization]{Inferring topologies via driving-based generalized  synchronization of two-layer networks}
%

\author{Yingfei~Wang$^{1}$ \& Xiaoqun~Wu$^{1,*}$ Hui~Feng$^{1,*}$ \&Jun-an Lu$^1$ \& Yuhua~Xu$^{2}$}

\address{$^1$School of Mathematics and Statistics, Wuhan University, Wuhan 430072, People's  Republic of China\\
$^2$School of Finance, Nanjing Audit University, Nanjing 211815,  People's  Republic of China}
\ead{xqwu@whu.edu.cn(Xiaoqun~Wu), hfeng.math@whu.edu.cn(Hui~Feng)}
\vspace{10pt}
\begin{indented}
\item[]December 2015
\end{indented}

%
%
\begin{abstract}
  The interaction topology among the constituents of a complex network plays a crucial role in the network's evolutionary mechanisms and functional behaviors.  However, some network topologies are usually  unknown or uncertain. Meanwhile, coupling delay are ubiquitous in various man-made and natural networks. Hence, it is necessary  to gain  knowledge of the whole or partial topology of a complex dynamical network by taking into consideration  communication delay.
In this paper, topology identification of complex dynamical networks   is investigated via generalized   synchronization of a two-layer network. Particularly, based on the LaSalle-type invariance principle of stochastic differential delay equations, an adaptive control technique  is proposed by constructing an auxiliary layer and designing proper control input and updating laws so that the unknown topology can be recovered upon successful generalized synchronization.   Numerical simulations are provided to illustrate the effectiveness of the proposed  method.  {\color{blue}The technique  provides a certain theoretical basis for topology inference of complex networks. In particular,} when the considered network is composed of systems with high-dimension or complicated dynamics, a simpler  response layer can be  constructed, which is conducive to   circuit  design.  Moreover,  it is practical  to take into consideration perturbations  caused by  control input.  Finally, the method is  applicable to infer topology of a  subnetwork embedded within  a complex system and locate hidden sources. {\color{blue} We hope the results can provide  basic insight into further research endeavors on understanding practical and economical topology inference of networks}.
\end{abstract}

%
%
%
%
%

\section{Introduction}\label{Intro}
Since the groundbreaking works of Watts and Strogatz \cite{watts1998collective} on small-world networks in 1998 and of Barab\'{a}si and Albert \cite{barabasi1999emergence} on scale-free networks in 1999, complex dynamical networks {as a new scientific branch have experienced rapid development and have permeated various fields}, such as mathematics, computer sciences, engineering sciences and so on \cite{liu2014multiobjective, sheshbolouki2015role}.
Initial research attention  {is mainly focused on complex dynamical networks'}  statistical properties and dynamical behaviors with previously known topologies. People rarely concentrate on inferring   connection topologies of networks,  partly because it involves the challenging inverse problem. It is worth noting that  topological structures of complex networks play a crucial role in determining their evolutionary mechanisms and functional behaviors \cite{newman2003structure, boccaletti2006complex}.
 Therefore,  it is of necessity and importance to gain knowledge of   unknown or uncertain topological structures of complex dynamical networks  {\color{blue}and  provide some theoretical guidance for topology identification of networks.}

In the past few years, a great variety  of methods have been developed for inferring network topologies,  using technologies such as {the synchronization-based method} \cite{zhou2007topology, wu2008synchronization, liu2009structure, zhao2010topology, wu2013structure}, compressive sensing \cite{wang2011predicting}, Bayesian estimation \cite{jansen2003bayesian}, recurrence \cite{marwan2007recurrence, romano2007estimation}, Granger causality test \cite{wu2011detecting, wu2012inferring}, node knockout \cite{nabi2012network}, echo state mechanism \cite{Li2012chaotic}, and so on. {Among  these}  methods, the one based on synchronization in which some adaptive controllers are designed so that an auxiliary response network can synchronize {with} the uncertain network and the topological parameters can be estimated simultaneously, has been paid wide attention to. It is worth mentioning {that  the synchronization } between the two networks   is   complete outer synchronization \cite{li2007synchronization, tang2008adaptive, wu2012outer}, which means that each pair of  nodes in the drive  and response networks  manifest completely identical dynamics upon outer synchronization.  However, nodes in different networks usually have diverse  dynamics, but the networks can still reach harmonious coexistence. This kind of synchronization is called {generalized outer  synchronization} \cite{wu2009generalized}, which  can be regarded as a special type of generalized synchronization and represents another degree of coherence \cite{kocarev1996generalized, juan2008generalized, senthilkumar2008transition, koronovskii2012generalized}. A typical example of the {generalized outer  synchronization}  is the predator and prey networks. Though individuals in predator network and prey network behave in quite different ways, they may finally coexist in harmony \cite{wu2009generalized}.  Therefore, inferring topological structures of complex networks via generalized synchronization is of practical and theoretical significance.

There are many factors influencing   network dynamics, such as stochastic perturbations and coupling delay. {It is noted  that a considerable number  of  existing works about topology inference  focus on  networks  free of noise perturbations. However, in real-world complex networks, noise is omnipresent. The determinant dynamical network is not real in practice since it usually omit some unknown factors in modeling.  For example, the signals transmitted between subsystems of a complex dynamical network are unavoidably subject to stochastic perturbations from environment, which may cause the loss of information contained in these signals \cite{yang2009stochastic}. The stochastic uncertainties can have   great impact on   behaviors of complex networks, such as in genetic oscillator networks, the gene regulation is subject to intracellular and extracellular noise perturbations and environmental fluctuation, so   cellular noise  will undoubtedly affect the dynamics of the networks quantitatively and qualitatively \cite{paulsson2004summing, raser2005noise, chen2014fuzzy}.} Besides, time delay, which is usually caused by finite signal transmission or memory effects,  frequently arises  in many physical and biological systems such as communication networks, neural networks and so on \cite{guo2014systematic, cao2005global, wei2015counterpart}.  {Therefore, to investigate and simulate more realistic networks, it is more practical and necessary to take into consideration  stochastic perturbations and coupling delay.}  Moreover,  {people are usually interested in part of a complex dynamical network.} For example,  for  multi-layer networks \cite{gomez2013diffusion}, one may just want to know the information of one layer of the whole {system}.  From the viewpoint of applicability, inferring the connection pattern of a subnetwork embedded within {a complex network} is a meaningful and necessary work.

Motivated by  above discussions, this paper {\color{blue}intends to give some theoretical basis and supports for topology identification  of practical complex networks. Particularly,}   topology inference of complex dynamical networks with   coupling delay {\color{blue}is investigated} via driving-based generalized  synchronization in a two-layer network. {\color{blue}The} interested network is considered as a drive layer, {an auxiliary network is constructed as the response layer.
 In addition, stochastic perturbations caused by control input from the drive layer to the response layer due to information transmission as well as circuit design
 are also taken into consideration.} Some control inputs and updating laws are designed so that  nodes in  the response layer  can reach {generalized outer synchronization} with their counterparts in the  drive layer, and the unknown topology of the drive layer can be adaptively inferred, {both in the sense of mean square}.

The rest of the paper is organized as follows. In Section \ref{sec:mod},  the network model and some preliminary lemmas  are introduced.  In Section \ref{sec:res}, by means of adaptive control, several   criteria for inferring topologies of complex dynamical networks with coupling delay are derived based on  the  LaSalle-type invariance principle of stochastic differential delay equations. In Section \ref{sec:sim}, numerical examples are examined to show the effectiveness of the identification method in the presence of {noise perturbations caused by control input and related factors}. Furthermore,  the method is applicable to recover partial topology of a network and detect hidden sources.   Finally, some conclusions and discussions are given in Section \ref{sec:con}.

\section{Preliminaries and network models}\label{sec:mod}
 Some necessary notations are first introduced. $\mathbb{R}^n$ and $\mathbb{R}^{n \times m}$ denote the $n$-dimensional Euclidean space and the set of all $n\times m$-dimensional real matrices, respectively. $\varotimes$ represents the Kronecker product, and the superscript $\top$ stands for the transpose of a vector or a matrix. $\|\cdot\|$ represents an arbitrary norm.   $\lambda_{\mbox{max}}(A)$ is the maximum eigenvalue of matrix $A$. $(\Omega, \mathcal{F}, \{\mathcal{F}_t\}_{t\geq 0}, \mathcal{P})$ is a complete probability space with a filtration $\{\mathcal{F}_t\}_{t\geq 0}$ satisfying  right continuity and $\mathcal{F}_0$ contains all $\mathcal {P}$-null sets. $E\{\cdot\}$ denotes the mathematical expectation. {$\dot{x}(t)$ represents the total derivative of $x$ with respect to $t$.}

Consider the following general complex dynamics network model with coupling delay,
\begin{equation}\label{eq:drive}
d\mathbf{x}_i(t)=\big(\mathbf{f}_i(t,\mathbf{x}_i(t))+\sum_{j=1}^Na_{ij}H\mathbf{x}_j(t-\tau)\big)dt,\,\,\,i=1,2,\cdots,N,
\end{equation}
where $\mathbf{x}_i = (x_{i1},x_{i2},\cdots,x_{im})^T \in \mathbb{R}^m$ is the state vector of the $i$-th node. $\mathbf{f}_i:\mathbb{R}_+ \times \mathbb{R}^m\rightarrow\mathbb{R}^m$ is a  smooth nonlinear vector-valued function which determines the intrinsic dynamical behavior of the $i$-th node in network (\ref{eq:drive}). $H\in \mathbb{R} ^{m\times m}$ is the inner coupling matrix linking interacted  component variables, and $A=(a_{ij})_{N\times N}\in \mathbb{R}^{N\times N}$ is the unknown or uncertain coupling configuration matrix representing the network's topological structure, whose entries are defined as follows: {if there is a link from node $j$ to node $i$, then $a_{ij}\neq 0$, otherwise, $a_{ij}= 0  (i\neq j)$, and  $a_{ii}=-\sum\limits_{j=1,j\neq i}^{N} a_{ij},(i=1,2,\cdots,N)$.} $\tau$ represents the information transmission delay between connected nodes.

It is  worth-noting that  network topologies play a pivotal  role in determining the emergence of collective behaviors and  governing the main features of relevant processes that take place in complex networks. Therefore,  it is of necessity and importance to gain knowledge of the intrinsic topology, which is usually unknown in many practical situations. In  network (\ref{eq:drive}), the node dynamics $\mathbf{f}_i$, the coupling matrix  $H$, and the communication delay  $\tau$ between nodes  are supposed to be known. Our purpose is to infer the  unknown topological matrix $A$ based on    signals  output from the considered network, which  will probably be contaminated by noise. For this purpose,  some necessary concepts and lemmas of stochastic differential equations are presented.

Consider  a nonautonomous $n$-dimensional stochastic differential delay equation
\begin{equation}\label{eq:delay stochastic equation}
d\mathbf{x}(t)=\mathbf{f}(t,\mathbf{x}(t),\mathbf{x}(t-\tau))dt+\mathbf{g}(t,\mathbf{x}(t), \mathbf{x}(t-\tau))d\mathbf{w}(t)
\end{equation}
on $t\geq 0$ with initial value $\xi \in C_{\mathcal{F}_0}^b([-\tau,0]; \mathbb{R}^n)$, where $C_{\mathcal{F}_0}^b([-\tau,0]; \mathbb{R}^n)$ denotes the family of all $\mathcal {F}_0$-measurable bounded $C([-\tau,0]; \mathbb{R}^n)$-valued random variables, the measurable functions $\mathbf{f}$ and $\mathbf{g}: \mathbb{R}_+ \times \mathbb{R}^n \times \mathbb{R}^n \rightarrow \mathbb{R}^n$ satisfy the local Lipschitz condition and the linear growth condition {\cite{mao2002note}}. Then, for any initial value $\xi \in C_{\mathcal{F}_0}^b([-\tau,0]; \mathbb{R}^n)$, Eq.(\ref{eq:delay stochastic equation}) has a unique solution denoted by $\mathbf{x}(t;\xi)$ on $t \geq -\tau$. Moreover, assume $\mathbf{f}(t,0,0)=\mathbf{g}(t,0,0)=0$, then Eq.(\ref{eq:delay stochastic equation}) admits a trivial solution $\mathbf{x}(t,0)\equiv 0$.

Let $C^{1,2}(\mathbb{R}_+ \times \mathbb{R}^n; \mathbb{R}_+)$ denote the family of all nonnegative functions $V(t,\mathbf{x})$ on $\mathbb{R}_+ \times \mathbb{R}^n$ which are continuously twice differentiable in $\mathbf{x}$ and once differentiable in $t$. Then  the diffusion operator $\mathcal {L} $ acting on $V(t,\mathbf{x})$ is
\begin{equation}\label{eq:duffusion operator}
\mathcal{L}V=V_t(t,\mathbf{x})+V_x(t,\mathbf{x})\mathbf{f}(t,\mathbf{x})
 +\frac{1}{2}\mbox{trace}[\mathbf{g}^T(t,\mathbf{x})V_{xx}(t,\mathbf{x})\mathbf{g}(t,\mathbf{x})],
\end{equation}
where $V_t(t,\mathbf{x})=\frac{\partial V(t,\mathbf{x})}{\partial t}$, $V_x(t,\mathbf{x})=\left(\frac{\partial V(t,\mathbf{x})}{\partial \mathbf{x}_1}, \frac{\partial V(t,\mathbf{x})}{\partial \mathbf{x}_2}, \cdots, \frac{\partial V(t,\mathbf{x})}{\partial \mathbf{x}_n}\right)$, $V_{xx}(t,\mathbf{x})=\left(\frac{\partial ^2 V(t,\mathbf{x})}{\partial \mathbf{x}_i\partial \mathbf{x}_j}\right)_{n\times n}$.

\begin{lem}\label{lem:stochastoc lasalle}
\cite{mao2002note} Assume that there is a function $V \in C^{1,2}(\mathbb{R}_+ \times \mathbb{R}^n; \mathbb{R}_+)$, a function $\gamma \in L^1(\mathbb{R}_+ ; \mathbb{R}_+)$ and continuous functions $\omega_1, \omega_2: \mathbb{R}^n\rightarrow  \mathbb{R}_+$ such that
\begin{equation*}
\mathcal {L} V(t,\mathbf{x},\mathbf{y})\leq \gamma(t) - \omega_1(\mathbf{x}) +\omega_2(\mathbf{y}),\,\,\, (t,\mathbf{x},\mathbf{y}) \in \mathbb{R}_+ \times \mathbb{R}^n \times \mathbb{R}^n,
\end{equation*}
\begin{equation*}
\omega_1(\mathbf{x}) > \omega_2(\mathbf{x}),\,\,\,\forall \mathbf{x}\neq 0,
\end{equation*}
\begin{equation*}
\lim_{\parallel \mathbf{x} \parallel \rightarrow \infty}  \inf_{0 \leq t < \infty} V(t,\mathbf{x})=\infty.
\end{equation*}
Then
\begin{equation}\label{eq:lasalle 1}
\lim_{t \rightarrow \infty} \mathbf{x}(t;\xi)=0 \,\,\, a.s.
\end{equation}
for any initial value $\xi \in C_{\mathcal{F}_0}^b([-\tau,0]; \mathbb{R}^n)$.
\end{lem}

\begin{lem}\label{lem:pq}
\cite{lu2007synchronization} For any vectors $\mathbf{x},\mathbf{y} \in \mathbb{R}^n$ and a {symmetric} positive definite matrix $Q\in \mathbb{R}^{n\times n}$, the following inequality holds: $2\mathbf{x}^T\mathbf{y}\leq \mathbf{x}^T Q \mathbf{x}+ \mathbf{y}^T Q^{-1}\mathbf{y}$.
\end{lem}

\section{Results}\label{sec:res}

To infer the unknown topology of network (\ref{eq:drive}), an auxiliary system consisting of an identical number of nodes as that of network (\ref{eq:drive}) is constructed as follows:
\begin{equation}\label{eq:response}
d\mathbf{y}_i(t)=\big(\mathbf{g}_i(t,\mathbf{y}_i(t))+\mathbf{u}_i\big)dt
+\psi_i(t)d\mathbf{w}_i(t),  ~~i=1,2,\cdots,N,
\end{equation}
where $\mathbf{y}_i = (x_{i1},x_{i2},\cdots,x_{in})^T \in \mathbb{R}^n$ is the state vector of the $i$-th node, $\mathbf{g}_i:\mathbb{R}_+ \times \mathbb{R}^n\rightarrow\mathbb{R}^n$ is a  smooth nonlinear vector-valued function governing the dynamical behavior of the $i$-th node. Furthermore, $\mathbf{u}_i$ is the control input to be designed for the $i$-th node. {The noise term is utilized to describe perturbations  caused by the control input, where information collected  from the drive layer will be  put into the response layer and thus noise will emerge due to information transmission and inaccurate circuit design.} Specifically, $\mathbf{w}_i(t)$ is an $n$-dimensional Brownian motion defined on the probability space  $(\Omega, \mathcal{F}, \{\mathcal{F}_t\}_{t\geq 0}, \mathcal{P})$ with
\begin{equation*}
\fl
E\{\mathbf{w}_i(t)\}=0,{\color{blue}E\{\mathbf{w}_i^2(t)\}=t}, E\{\mathbf{w}_i(s)\mathbf{w}_i(t)\}=\min\{s,t\}(s\neq t), E\{\mathbf{w}_i(t)\mathbf{w}_j(t)\}=0(i\neq j),
\end{equation*}
$\psi_i(t)$ is the vector-form noise intensity function which describes the intensity of uncertain perturbations to the $i$-th node.

Since each node in the auxiliary network (\ref{eq:response})  receives driving signals from its counterpart in the considered network (\ref{eq:drive}), the two networks form a two-layer network with one-to-one unidirectional connections. In the following,     network (\ref{eq:drive}) is regarded as the drive  layer, and
network (\ref{eq:response}) is regarded as the  response layer.  To infer  topologies of the considered network,
 controllers are designed so that the response layer can harmoniously oscillate with the drive layer and topological parameters of network (\ref{eq:drive})  are reconstructed. {In addition, due to the fact that node dynamics in the two layers are usually different, one will use generalized outer synchronization rather than complete outer synchronization to characterize the harmonious oscillation between the two layers.}

 Particularly, let $\mathbf{e}_i(t)=\mathbf{y}_i(t)-\phi_i(\mathbf{x}_i(t))$ denote  the generalized outer synchronization error between the $i$-th nodes in the two layers, where $\phi_i:\mathbb{R}^m\rightarrow \mathbb{R}^n$ is a {once} continuously differentiable vector function describing  functional relationships between corresponding nodes. The two network layers are said to achieve generalized outer synchronization (or one can say the two-layer network achieves generalized synchronization) {in the sense of mean square if
\begin{equation*}
\lim_{t\rightarrow\infty}E\|\be_i(t)\|^2=0,\,\,\,i=1,2,\cdots, N.
\end{equation*}
}
Let
\begin{equation}\label{eq:updating laws3}
\dot{\hat{a}}_{ij}(t)=-\delta_{ij}\be_i^T(t)D\ph_i(\bx_i)H\bx_j(t-\tu),\,\,\,i,j=1,2,\cdots,N
\end{equation}
be updating laws to track the unknown topological parameters $a_{ij}~(i,j=1,2,\cdots,N)$ in the drive layer.
  The control input $\mathbf{u}_i$  for the $i$-node in the response layer and the adaptive feedback gain are thus designed as
\begin{eqnarray}\label{eq:updating laws1}
\bu_i(t)= & D\phi_i(\bx_i)\mathbf{f}_i(t,\bx_i(t))-\mathbf{g}_i(t,\ph_i(\bx_i(t)))\nonumber\\
& +D\ph_i(\bx_i)\sum_{j=1}^N\hat{a}_{ij}H\bx_j(t-\tau) -d_i(t)\be_i(t),
\end{eqnarray}
and
\begin{equation}\label{eq:updating laws2}
\dot{d}_i(t)=k_i\be_i^T(t)\be_i(t),\,\,\,i=1,2,\cdots,N,
\end{equation}
respectively,
where $D\phi_i(\mathbf{x}_i)$ is the Jacobian matrix of $\phi_i(\mathbf{x}_i(t))$,   $\delta_{ij}>0$ and $k_i>0$ are arbitrary constants.

Let $\tilde{a}_{ij}=\hat{a}_{ij}-a_{ij}$, then the error dynamics between the   layers (\ref{eq:drive}) and (\ref{eq:response}) can be described as
\begin{eqnarray}\label{eq:error}
d\mathbf{e}_i(t) &=& d\by_i(t)-D\ph_i(\bx_i)d\bx_i(t) \nonumber\\
&=& \big(\mathbf{g}_i(t,\by_i(t))-D\ph_i(\bx_i)\mathbf{f}_i(t,\bx_i(t))
 -D\ph_i(\bx_i)\sum_{j=1}^Na_{ij}H\bx_j(t-\tau)\nonumber\\
 && +\bu_i\big)dt
+\p_i(t)d\mathbf{w}_i(t) \nonumber\\
 &=& \big(\mathbf{g}_i(t,\by_i(t))-\mathbf{g}_i(t,\ph_i(\bx_i(t)))
+D\ph_i(\bx_i)\sum_{j=1}^N\tilde{a}_{ij}H\bx_j(t-\tu) \nonumber\\
&& -d_i(t)\be_i(t)\big)dt
 + \p_i(t)d\mathbf{w}_i(t),\,\,\,i=1,2,\cdots,N,
\end{eqnarray}

\begin{asp}\label{asp:lipuxizi}
 {(Lipschitz condition)}For the nonlinear vector function  $\mathbf{g}_i(\cdot)$, there exists  a positive constant $\alpha_i$ such that for
$i=1,2,\cdots,N$,
\begin{equation*}
\|\mathbf{g}_i(t,\mathbf{y}(t))-\mathbf{g}_i(t,\mathbf{x}(t))\| \leq \alpha_i \|\mathbf{y}(t)-\mathbf{x}(t)\|
\end{equation*}
holds for any $\mathbf{x}(t),\mathbf{y}(t)\in \mathbb{R}^n$.
\end{asp}

\begin{asp}\label{asp:noise bound}
 Suppose that {$\psi_i(t)$} is bounded, and there exist nonnegative constants $\mu_i,\nu_i\geq 0 $ such that for
$i=1,2,\cdots,N$,
$$
\mbox{trace}(\p_i^T(t)\p_i(t)) \leq 2\mu_i\be_i^T(t)\be_i(t)+2\nu_i\be_i^T(t-\tu)\be_i(t-\tu).
$$
\end{asp}

{
\begin{rmk}\label{rmk:noise bound}
For the purpose of inferring the unknown topology of the drive layer, a response layer is constructed.
  The noise term in the constructed auxiliary network  is utilized to describe perturbations  caused by the control input, where information collected  from the drive layer (such as $\mathbf{x}_i$) will be  put into the response layer and thus noise will emerge due to information transmission as well as inaccurate circuit design. During this process, the noise is inevitably influenced by both the input signal $\mathbf{x}_i$ and response dynamics $\mathbf{y}_i$.  This is a commonly-employed condition in many literatures \cite{lu2012synchronization, yang2010adaptive}.
\end{rmk}
}

\begin{asp}\label{asp:independent}
 $\{D\ph_i(\bx_i)H\bx_i(t)\}_{i=1}^N$ are linearly independent on the orbit $\{\bx_i(t)\}_{i=1}^N$ of the  {generalized outer synchronization} manifold $\{\by_i(t)=\ph_i(\bx_i(t))\}_{i=1}^N$ between the two network layers.
\end{asp}

With these assumptions, the main topology identification result based on generalized  synchronization in a two-layer network is given as follows.

\begin{thm}\label{thm:main result}
 Let Assumptions \ref{asp:lipuxizi}-\ref{asp:independent} hold. Then the uncertain coupling configuration matrix $A$ of the considered  network (\ref{eq:drive}) can be identified by the estimated values $\hat{A}$ with probability one via adaptive controllers and updating laws (\ref{eq:updating laws3})-(\ref{eq:updating laws2}). Meanwhile, the response layer (\ref{eq:response}) reaches {generalized outer synchronization} with the drive layer  (\ref{eq:drive}) {in   mean square}.
\end{thm}

\textbf{Proof.} Consider the following Lyapunov function $V\in C^{1,2}(\mathbb{R}_+ \times \mathbb{R}^{nN+N^2+N}; \mathbb{R}_+)$,
\begin{equation}\label{eq:lyapunov function}
\fl
V(t;\mathbf{e}_i,\tilde{a}_{ij},d_i)= \frac{1}{2}\sum_{i=1}^N\mathbf{e}_i^T(t)\mathbf{e}_i(t)+
\frac{1}{2}\sum_{i=1}^N\sum_{j=1}^N \frac{1}{\delta_{ij}}\tilde{a}_{ij}^2+
\frac{1}{2}\sum_{i=1}^N \frac{1}{k_i}(d_i(t)-d_i^*)^2,
\end{equation}
where $d_i^*~(i=1,2,\cdots,N)$ are positive constants to be determined.

The diffusion operator $\mathcal {L} $ defined in (\ref{eq:duffusion operator}) onto the function $V$ along the trajectories of the error dynamics (\ref{eq:error})  gives
\begin{eqnarray*}
\fl
\mathcal {L}V =
\sum_{i=1}^N\be_i^T(t)\big(\mathbf{g}_i(t,\by_i(t))-\mathbf{g}_i(t,\ph_i(\bx_i(t)))
 +D\ph_i(\bx_i)\sum_{j=1}^N\tilde{a}_{ij}H\bx_j(t-\tu)
 -d_i(t)\be_i(t)\big)\\
\fl
\qquad \quad +\sum_{i=1}^N\sum_{j=1}^N \frac{1}{\delta_{ij}}\tilde{a}_{ij}\dot{\hat{a}}_{ij}
+\sum_{i=1}^N \frac{1}{k_i}(d_i(t)-d_i^*)\dot{d}_i(t)+
\frac{1}{2}\sum_{i=1}^N \mbox{trace}(\p_i^T(t) \p_i(t)).
\end{eqnarray*}
Together with the updating laws (\ref{eq:updating laws3}) and (\ref{eq:updating laws2}), one gets
\begin{eqnarray*}
\fl
\mathcal {L}V = \sum_{i=1}^N\be_i^T(t)\big(\mathbf{g}_i(t,\by_i(t))-\mathbf{g}_i(t,\ph_i(\bx_i(t)))
 +D\ph_i(\bx_i)\sum_{j=1}^N\tilde{a}_{ij}H\bx_j(t-\tu)
 -d_i(t)\be_i(t)\big)\\
\fl
\qquad \quad -\sum_{i=1}^N\sum_{j=1}^N \tilde{a}_{ij}\be_i^T(t)D\ph_iH\bx_j(t-\tu)
 + \sum_{i=1}^N (d_i(t)-d_i^*)\be_i^T(t)\be_i(t)+
\frac{1}{2}\sum_{i=1}^N \mbox{trace}(\p_i^T(t) \p_i(t))\\
\fl
\qquad \!= \sum_{i=1}^N\be_i^T(t)\big(\mathbf{g}_i(t,\by_i(t))-\mathbf{g}_i(t,\ph_i(\bx_i(t)))\big)
 -\sum_{i=1}^N d_i^*\be_i^T(t)\be_i(t)+
\frac{1}{2}\sum_{i=1}^N \mbox{trace}(\p_i^T(t) \p_i(t)).
\end{eqnarray*}
Following from Assumptions \ref{asp:lipuxizi} and   \ref{asp:noise bound}, one has
\begin{eqnarray*}
 \mathcal {L}V &\leq &\sum_{i=1}^N (\alpha_i-d_i^*)\be_i^T(t)\be_i(t)
 +\sum_{i=1}^N \big( \mu_i\be_i^T(t)\be_i(t)+\nu_i\be_i^T(t-\tu)\be_i(t-\tu) \big) \\
 &\leq & \sum_{i=1}^N (\alpha_i+\mu_i-d_i^*)\be_i^T(t)\be_i(t) + \sum_{i=1}^N\nu_i\be_i^T(t-\tu)\be_i(t-\tu).
\end{eqnarray*}
Let $\mathbf{e}(t)=(\mathbf{e}_1(t)^T,\mathbf{e}_2(t)^T,\dots,\mathbf{e}_N(t)^T )^T \in \mathbb{R}^{nN}$, $\alpha=\max_{1\leq i \leq N}\{\alpha_i\}$, $d^*=\max_{1\leq i \leq N}\{d_i^*\}$, $\mu=\max_{1\leq i \leq N}\{\mu_i\}$, $\nu=\max_{1\leq i \leq N}\{\nu_i\}$, then
\begin{eqnarray*}
 \mathcal {L}V
 & \leq & -(d^*-\alpha-\mu)\be^T(t)\be(t)+ \nu\be_i^T(t-\tu)\be_i(t-\tu)\\
 &\triangleq & -\omega_1(\be(t))+\omega_2(\be(t-\tau)),
\end{eqnarray*}
where $\omega_1(\bx)\triangleq (d^*-\alpha-\mu)\mathbf{x}^T\mathbf{x}$, $\omega_2(\mathbf{x})\triangleq \nu\bx^T\bx$.

It is obvious that if $d^* > \alpha + \mu + \nu$, then $\omega_1(\bx) > \omega_2(\bx)$ for any $\bx \neq 0$. In addition, $\lim_{\|\be\|\rightarrow \infty}\inf_{0\leq t < \infty} V  = \infty $ and $\p_i$ is bounded.
 Following from Lemma \ref{lem:stochastoc lasalle}, one obtains $\lim_{t\rightarrow \infty} \be(t)=0$ a.s., which implies that {$\lim_{t\rightarrow\infty}E\|\be_i(t)\|^2=0$ and} the solutions regarding Eqs. (\ref{eq:updating laws3}), (\ref{eq:updating laws2}) and (\ref{eq:error}) starting from $ \mathbb{R}^{nN+N^2+N}$ will asymptotically stabilize at $M=\{(\mathbf{e}_i, \tilde{a}_{ij}, d_i)\in \mathbb{R}^{nN+N^2+N}: \mathbf{e}=0\}$ with probability one. In addition, together with  system  (\ref{eq:error}) and Assumption \ref{asp:independent}, one gets $M =\{ \mathbf{e}=0,\tilde{a}_{ij}=0, d_i=constants  \}$ for $ i,j=1,2,\dots,N $.  That is,  with the controllers and updating laws (\ref{eq:updating laws3})-(\ref{eq:updating laws2}), the response network layer (\ref{eq:response}) reaches  generalized outer synchronization with the drive layer (\ref{eq:drive}) and the unknown coupling configuration matrix $A$ is successfully identified by $\hat{A}$  {in the sense of mean square}. This completes the proof.

\begin{rmk}\label{rmk:synnetric and irreducible}
It is clear that the coupling configuration matrix $A$ is not  required to be {symmetric, irreducible or diffusive}, and the inner coupling matrix $H$ is not necessarily symmetric. This indicates  that the unknown structure of network (\ref{eq:drive})  can be undirected or directed, connected or disconnected. It may even contain isolated nodes or clusters. In addition, there is not any constraint imposed on the nodal or  coupling form in the constructed response network  layer (\ref{eq:response}). Moreover, each node in the two-layer network may have various dynamical behaviors. Therefore, the identification technique  is applicable to a large variety of complex dynamical networks with communication delay.
\end{rmk}

\begin{rmk}\label{rmk_general model}
Unlike many previous schemes \cite{yu2011inferring, zhou2007topology, wu2008synchronization, liu2009structure, zhao2010topology}, the constructed auxiliary network can be consisting of any kind of dynamical nodes other than  nodes with identical dynamics as  their counterparts in the drive layer. Therefore, if the considered network is composed of nodes carrying very complicated node dynamics or high node dimensions, one can design a response layer using nodes with much simpler dynamics. {Moreover, the topological structure of the auxiliary layer can be any form, which does not affects the inferring topology for the drive layer. Hence, the form of auxiliary system is more practical for circuit design.}
\end{rmk}

\begin{rmk}\label{rmk:linear independence condition}
It should be especially pointed  out that the linear independence condition in Assumption \ref{asp:independent} is a very essential condition for guaranteeing successful topology identification. {The linear independence condition is needed for  theoretical analysis. However, it is very difficult to verify this condition in practice.  Chen et al. declared that complete synchronization in the unknown network will make the linearly independence condition unsatisfied, then the identification of the unknown topology is impossible \cite{chen2009synchronization}. They also found that partial synchronization in the unknown network implies a part of topology being unidentifiable, and the results can be extended to projection synchronization and some generalized synchronization. In addition, Liu et al. declared that synchronization of the drive network is harmful to the identification of the topological structure as it is difficult to satisfy Assumption  3.3 \cite{liu2009structure}.  They also pointed out that it is often difficult to recover the topology of an unknown network with identical nodes, since the network easily reaches some kind of inner synchronization. Therefore, it is  practical to  check whether the nodes of the unknown network achieve synchronization instead of  checking the linear independence condition. Moreover, diverse dynamics and chaotic behaviors of the nodes within the unknown network can facilitate topology identification.}
\end{rmk}

Based on Theorem \ref{thm:main result}, one can easily derive the following corollaries.

\begin{cor}\label{cor:non coupling delay}
Suppose Assumptions \ref{asp:lipuxizi}-\ref{asp:independent} hold. If the considered network (\ref{eq:drive}) is free of coupling delay, that is,  $\tau=0$, then the uncertain coupling configuration matrix $A$ can be inferred by  $\hat{A}$ {in mean square}   via the controllers
\begin{equation}\label{eq:non delay laws1}
\fl
\bu_i(t)= D\phi_i(\bx_i)\mathbf{f}_i(t,\bx_i(t))-\mathbf{g}_i(t,\ph_i(\bx_i(t)))
+D\ph_i(\bx_i)\sum_{j=1}^N\hat{a}_{ij}H\bx_j(t)-d_i(t)\be_i(t),
\end{equation}
and updating laws
\begin{equation}\label{eq:non delay laws2}
\fl
\dot{d}_i(t)=k_i\be_i^T(t)\be_i(t),\,\,\,
\dot{\hat{a}}_{ij}(t)=-\delta_{ij}\be_i^T(t)D\ph_i(\bx_i)H\bx_j(t),
\,\,\,i,j=1,2,\cdots,N,
\end{equation}
where $k_i, \delta_{ij}~ (i,j=1,2,\cdots,N)$ are arbitrary positive constants.
\end{cor}

\begin{cor}\label{cor:non noise}
Suppose Assumptions \ref{asp:lipuxizi} and \ref{asp:independent} hold. If there is no noise perturbations, that is,  {$\p_i(t)=0,(i=1,2,\cdots,N)$}, then the uncertain coupling configuration matrix $A$ can be identified by the estimated values $\hat{A}$  via   updating laws and controllers (\ref{eq:updating laws3})-(\ref{eq:updating laws2}) upon successful generalized out synchronization in the two network layers ({\ref{eq:drive}}) and (\ref{eq:response}).
\end{cor}

\begin{cor}\label{cor:same dimension and dynamics}
Suppose Assumptions \ref{asp:lipuxizi}-\ref{asp:independent} hold. If the $i$-th node in the response layer (\ref{eq:response}) is designed to be the same as its counterpart in the drive layer (\ref{eq:drive}), that is,  $ \mathbf{f}_i(\cdot)=\mathbf{g}_i(\cdot)$, then the uncertain coupling configuration matrix $A$ can be identified by the estimated values $\hat{A}$ and  complete outer synchronization  between the two network layers {can be realized in mean square} via the  controllers
\begin{equation}\label{eq:same dimension and dynamics laws1}
\fl
\bu_i(t)=\sum_{j=1}^N\hat{a}_{ij}H\by_j(t-\tu)-d_i(t)\be_i(t),\,\,\,i=1,2,\cdots,N,
\end{equation}
and updating laws
\begin{equation}\label{eq:same dimension laws2}
\fl
\dot{d}_i(t)= \,\,k_i\be_i^T(t)\be_i(t),\,\,\,
\dot{\hat{a}}_{ij}(t)=-\delta_{ij}\be_i^T(t)D\ph_i(\bx_i)H\by_j(t-\tu),
\,\,\,i,j=1,2,\cdots,N,
\end{equation}
where $k_i, \delta_{ij} ~(i,j=1,2,\cdots,N)$ are arbitrary positive constants.
\end{cor}

Furthermore, if  the response layer (\ref{eq:response}) receives input signals from the drive layer (\ref{eq:drive}) free of noise perturbations and there is no information transmission delay between node pairs, then with controllers (\ref{eq:non delay laws1}) and (\ref{eq:non delay laws2}), the uncertain coupling configuration matrix $A$ can be identified by the estimated values $\hat{A}$ upon generalized outer synchronization between the two layers. This result covers the latest result presented in   \cite{zhang2014recovering}. Moreover, the auxiliary layer (\ref{eq:response}) and controllers (\ref{eq:non delay laws1}) are much simpler than their counterparts in \cite{zhang2014recovering}.     It is worth mentioning that the auxiliary network layer (\ref{eq:response})  constructed in this paper contains no  connections between its  nodes,  thus the auxiliary system and controllers  are much simpler and more practical than previous results  \cite{zhou2007topology, wu2008synchronization, liu2009structure, zhao2010topology},  \cite{zhang2014recovering}.

\section{Numerical simulations}\label{sec:sim}

\begin{figure*}[!t]
\centering
\includegraphics[height=5.0cm]{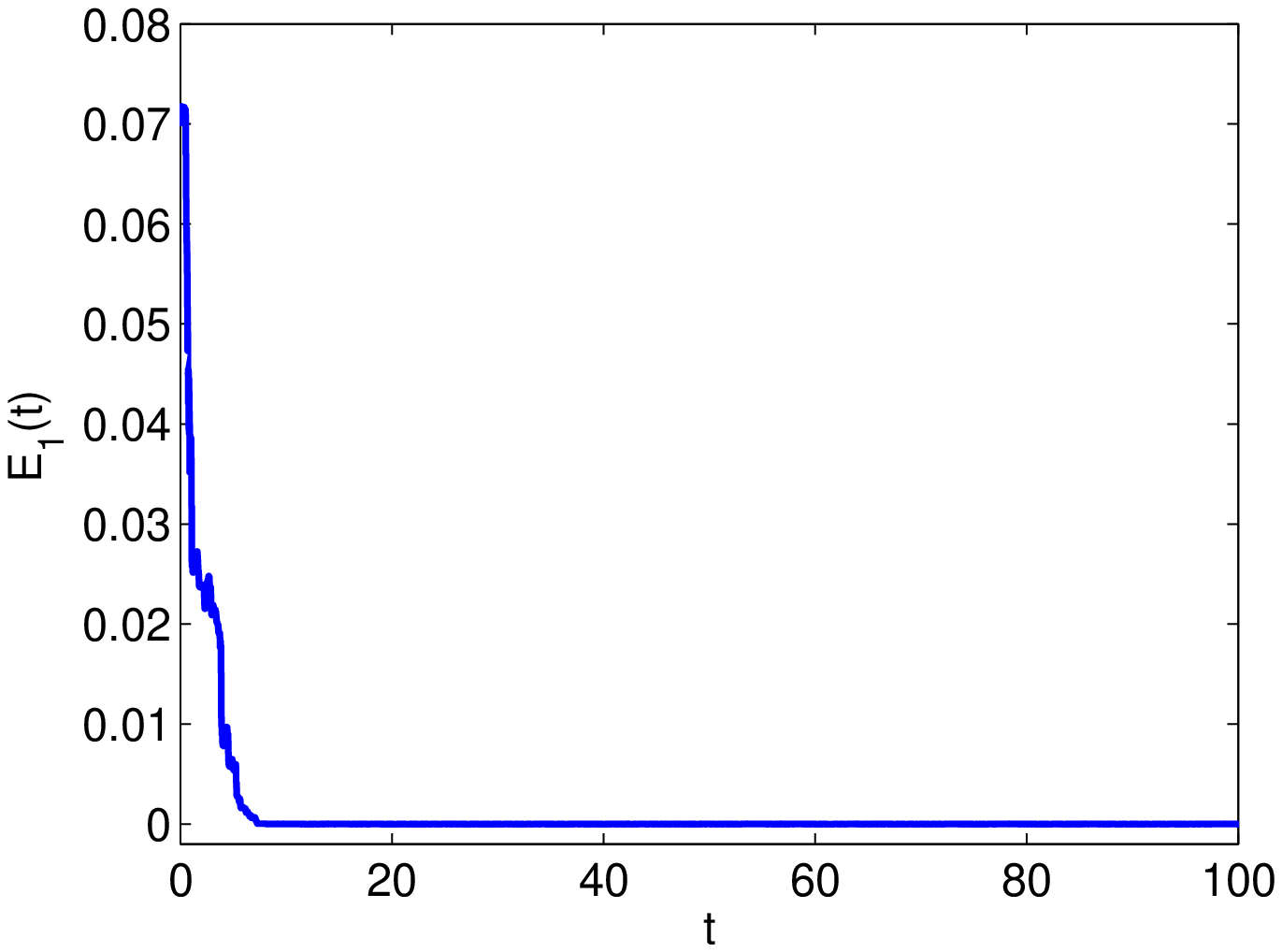}
\hspace*{1ex}
\includegraphics[height=5.0cm]{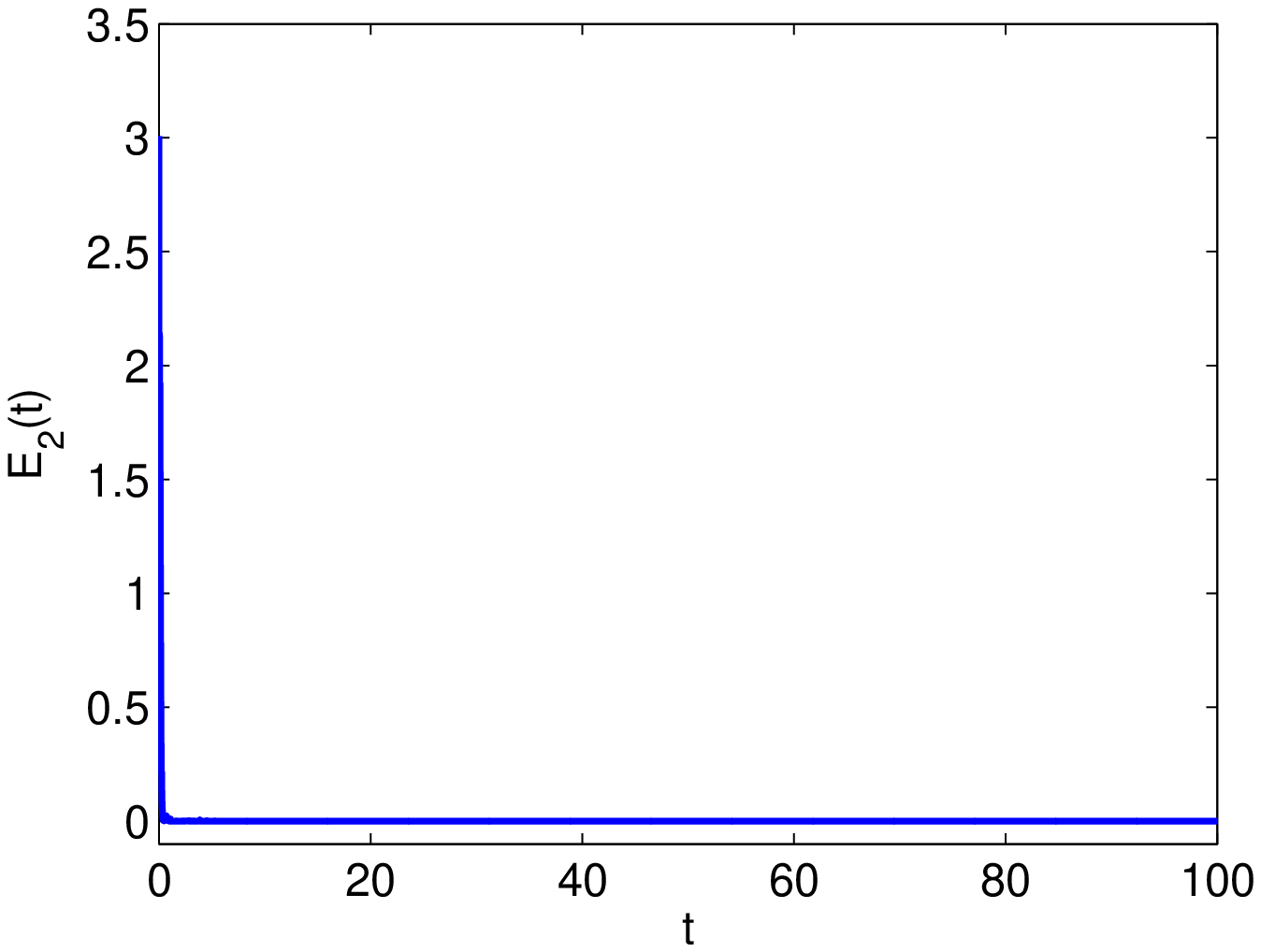}
\caption{Identification error (left) and synchronization error (right) of the two-layer L\"{u}-Chua network.}
\label{fig:err_2}

\vspace*{1.5ex}

\centering
\includegraphics[height=5.0cm]{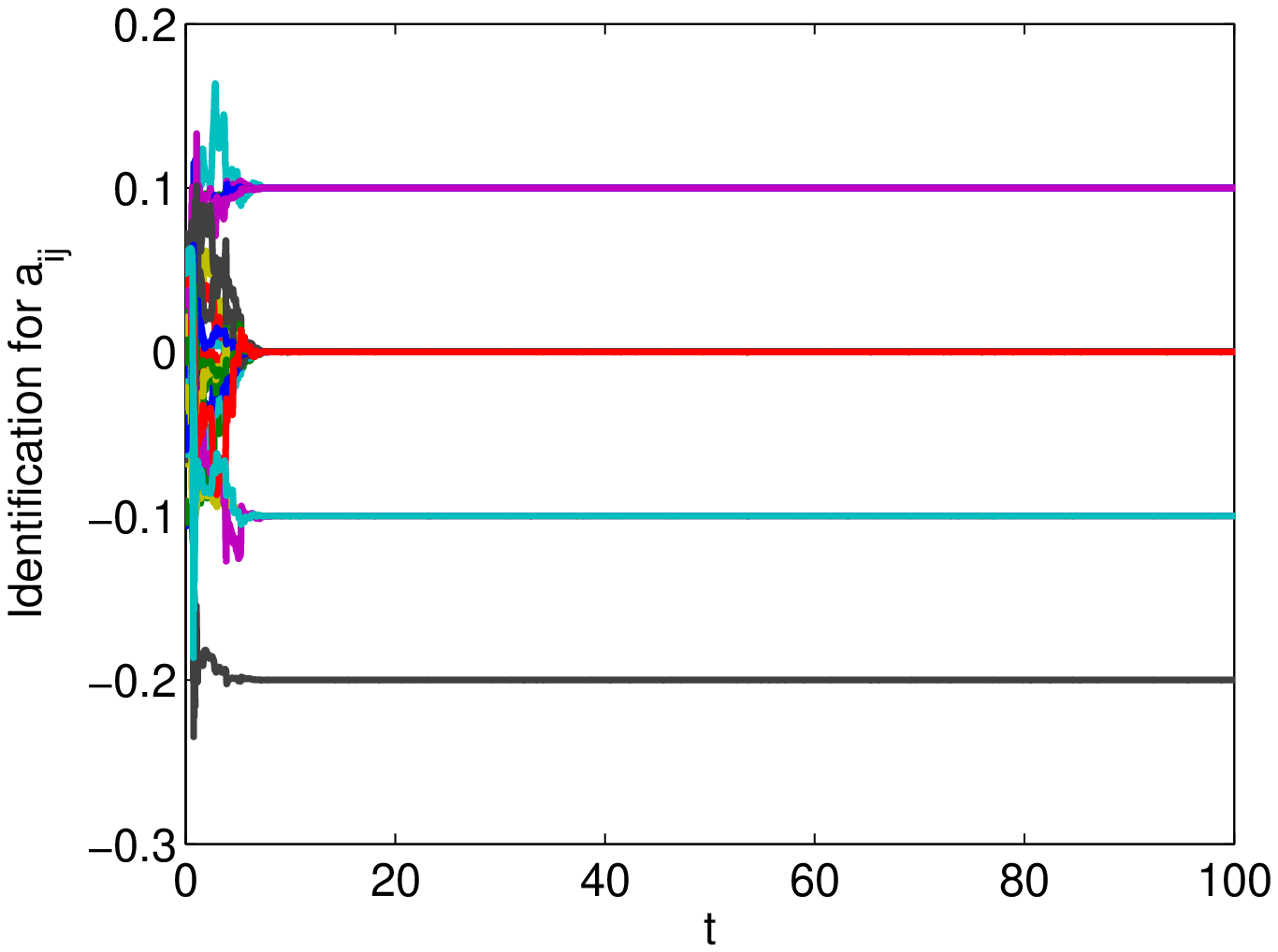}
\hspace*{1ex}
\includegraphics[height=5.0cm]{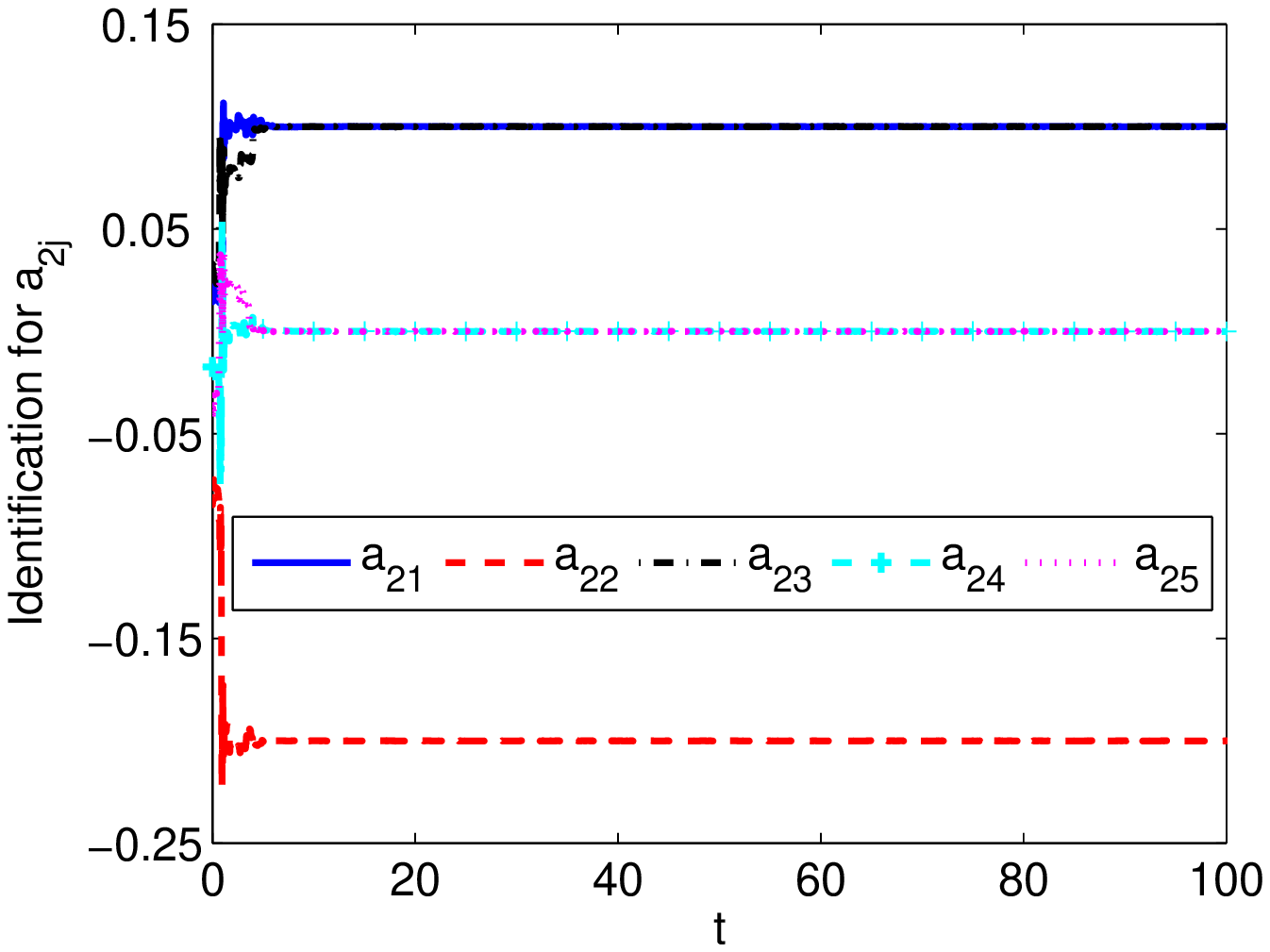}
\caption{Time evolution of $\hat{a}_{ij}~(i,j=1,2,...,5)$ (left) and  $\hat{a}_{2j}~(j=1,2,...,5)$ (right) of the two-layer L\"{u}-Chua network.}
\label{fig:aij_2}
\end{figure*}

\begin{figure*}[!t]
\centering
\includegraphics[height=5.0cm]{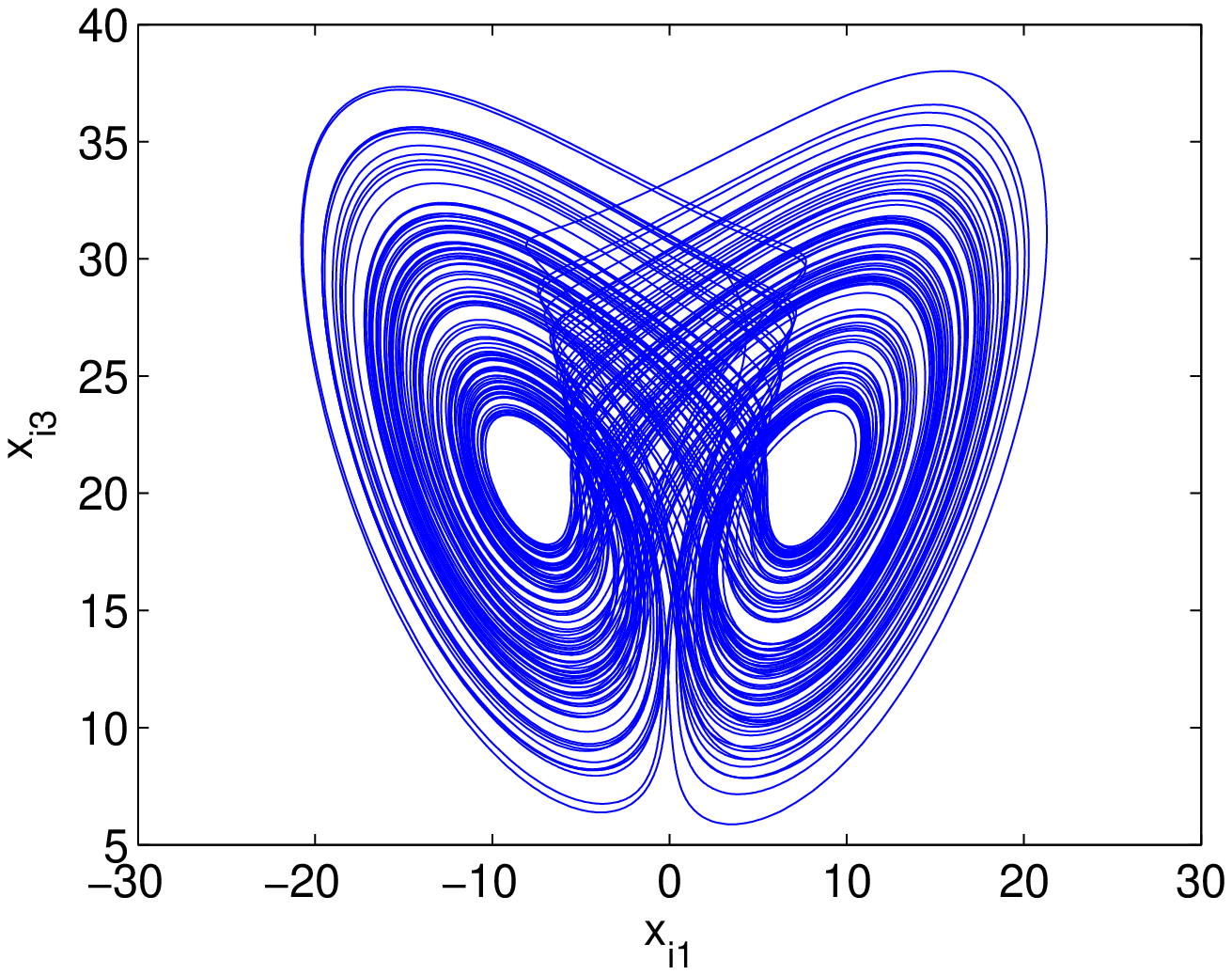}
\hspace*{1ex}
\includegraphics[height=5.0cm]{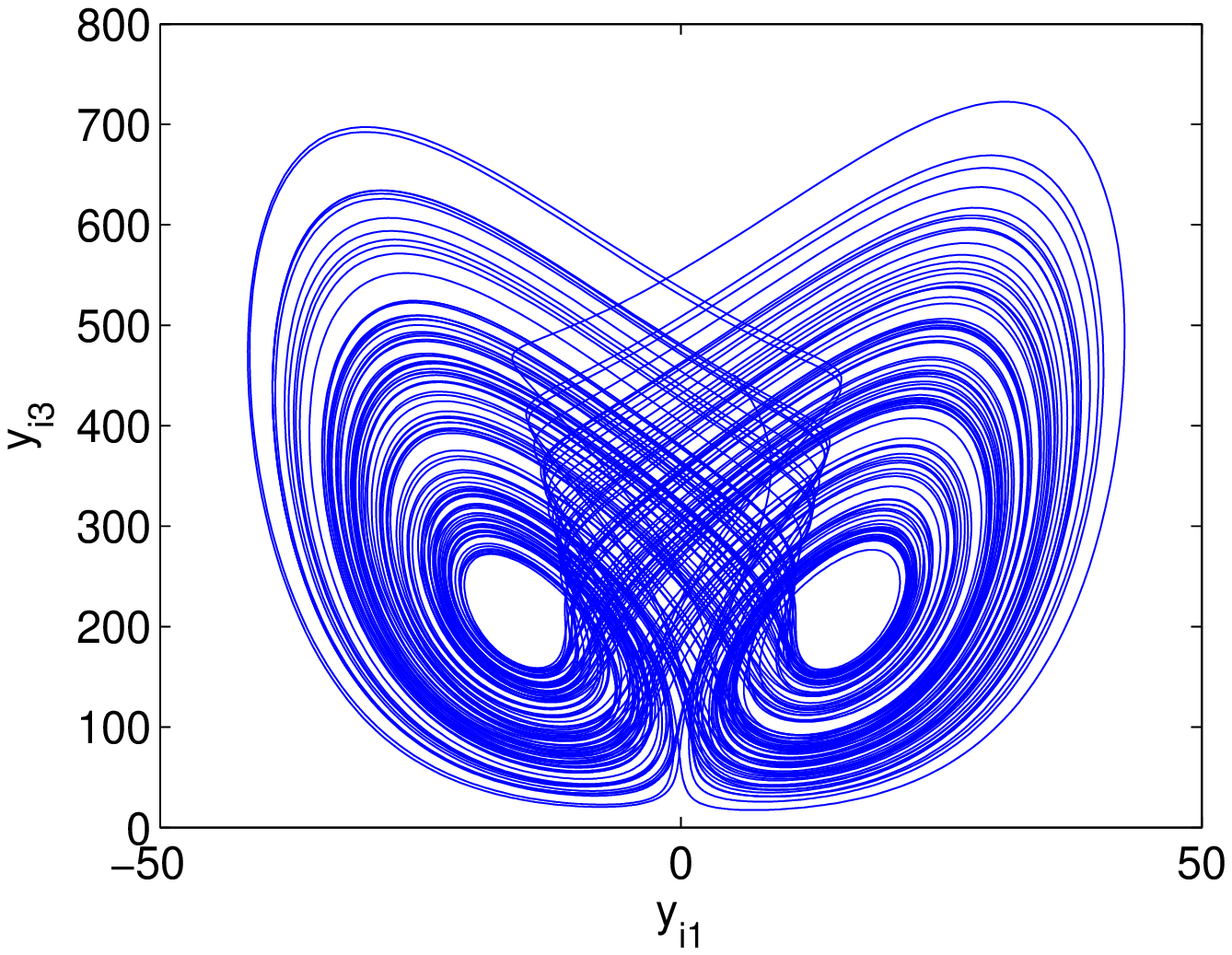}
\caption{Phase diagrams of node dynamics  in the  two-layer L\"{u}-Chua network. Left: projection  in the $(x_{i1},x_{i3})$-plane of  node 2 in the drive layer; right: projection in the $(y_{i1},y_{i3})$-plane of the node 2 in the response layer. }
\label{fig:xi1xi3_2_2}

\vspace*{1.5ex}

\centering
\includegraphics[height=5.0cm]{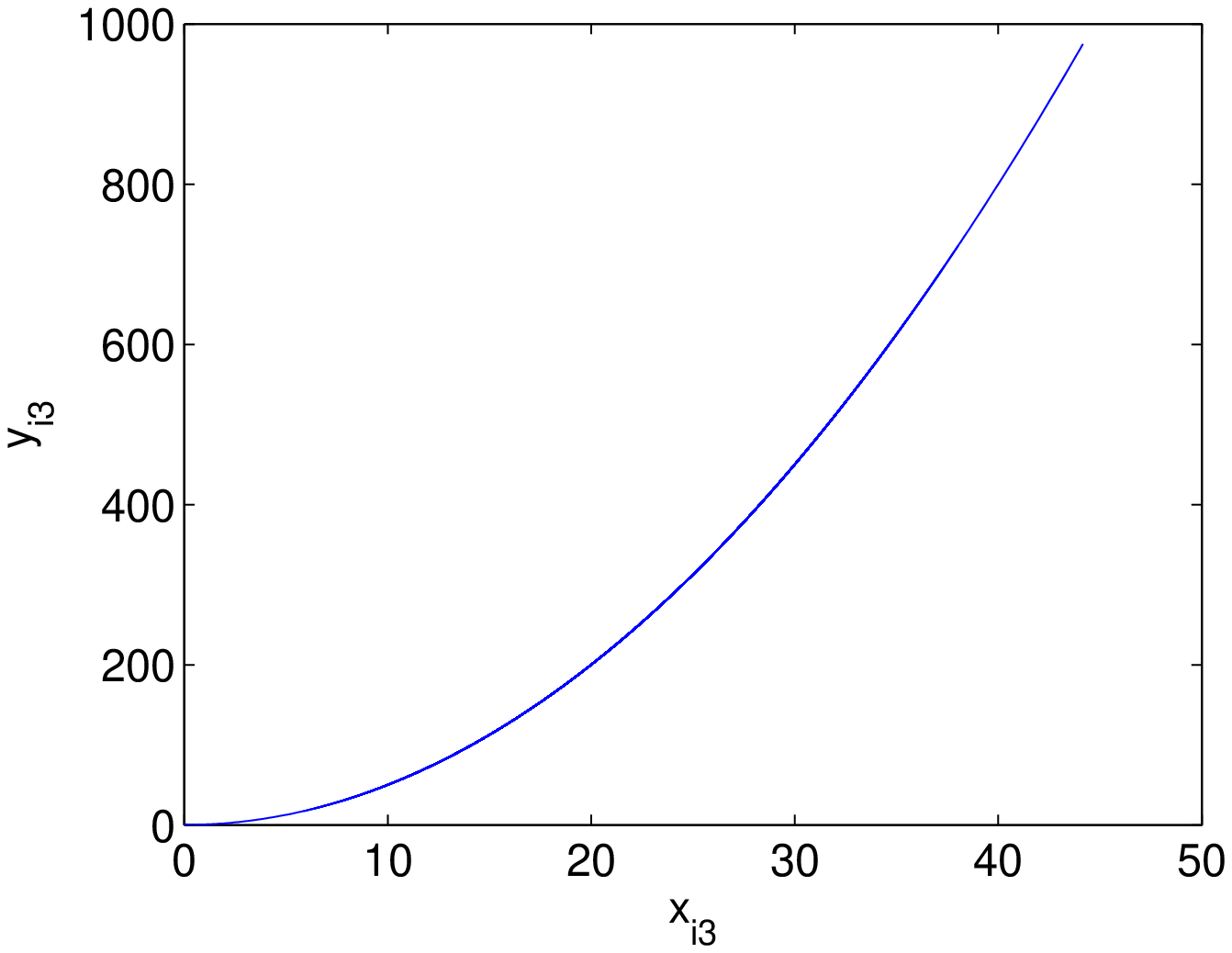}
\hspace*{1ex}
\includegraphics[height=5.0cm]{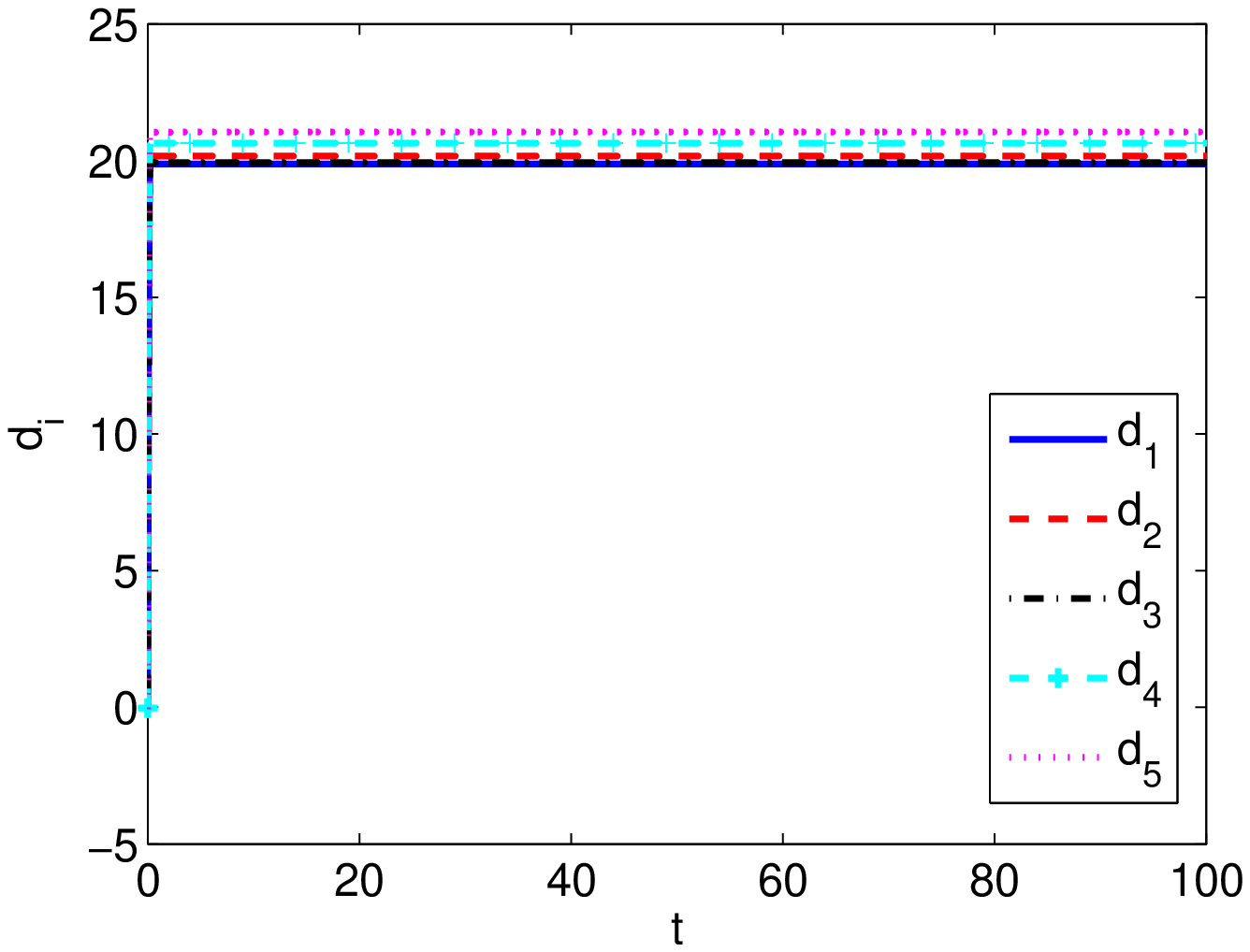}
\caption{Relationship between the component variables $x_{i3}$ and $y_{i3}$ of node 2 (left) and time evolution of adaptive feedback gains $d_i~(i=1,2,...,5)$ (right) of the two-layer L\"{u}-Chua network.}
\label{fig:di_2}
\end{figure*}

\begin{figure*}[!t]
\centering
\includegraphics[height=5.0cm]{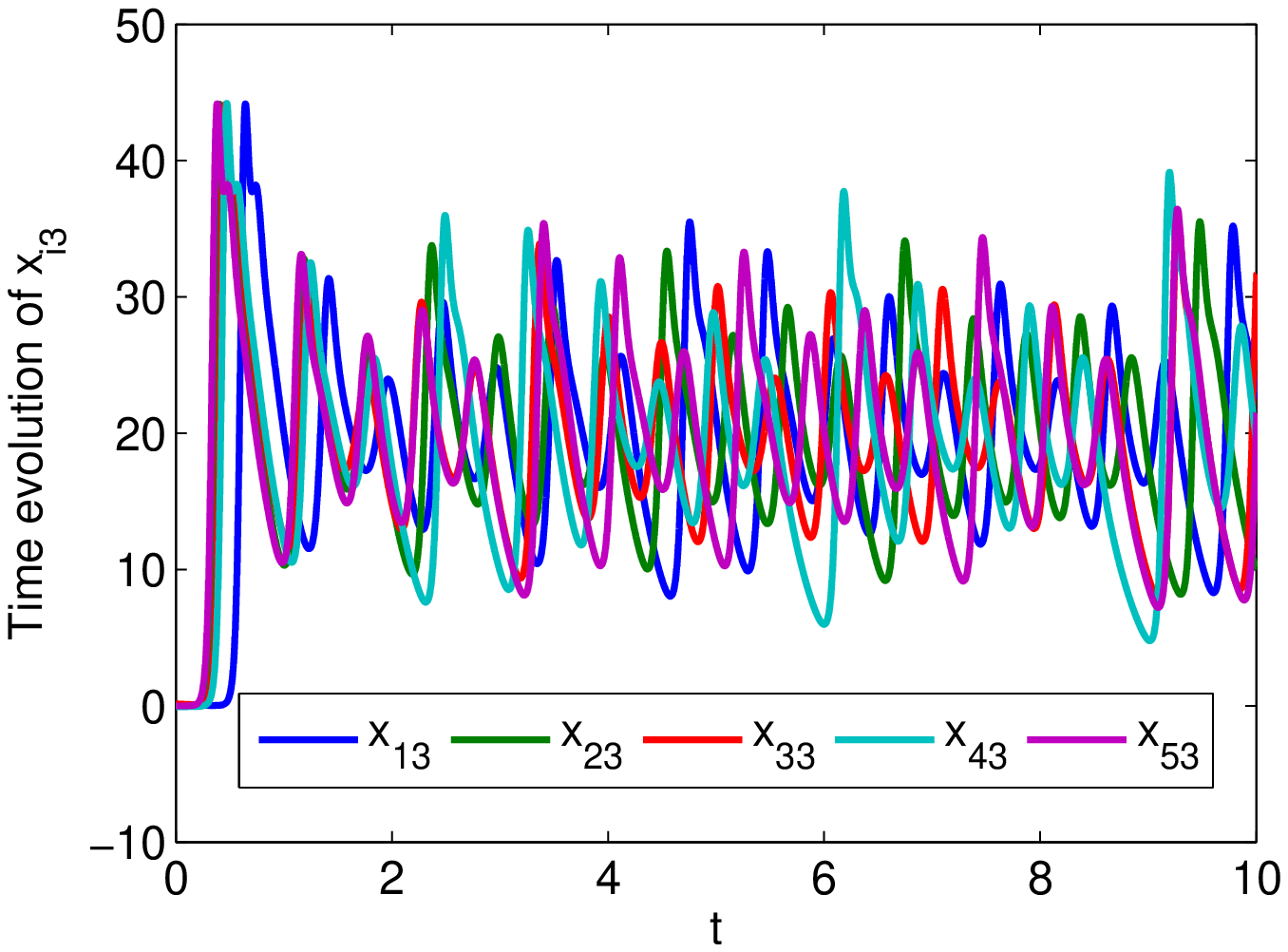}
\caption{Time evolution of the component variables $x_{i3}~(i=1,2,...,5)$ in the drive layer  of the two-layer  L\"{u}-Chua network with successful topology identification.}
\label{fig:ide_suc_xi3}

\vspace*{1.5ex}

\centering
\includegraphics[height=5.0cm]{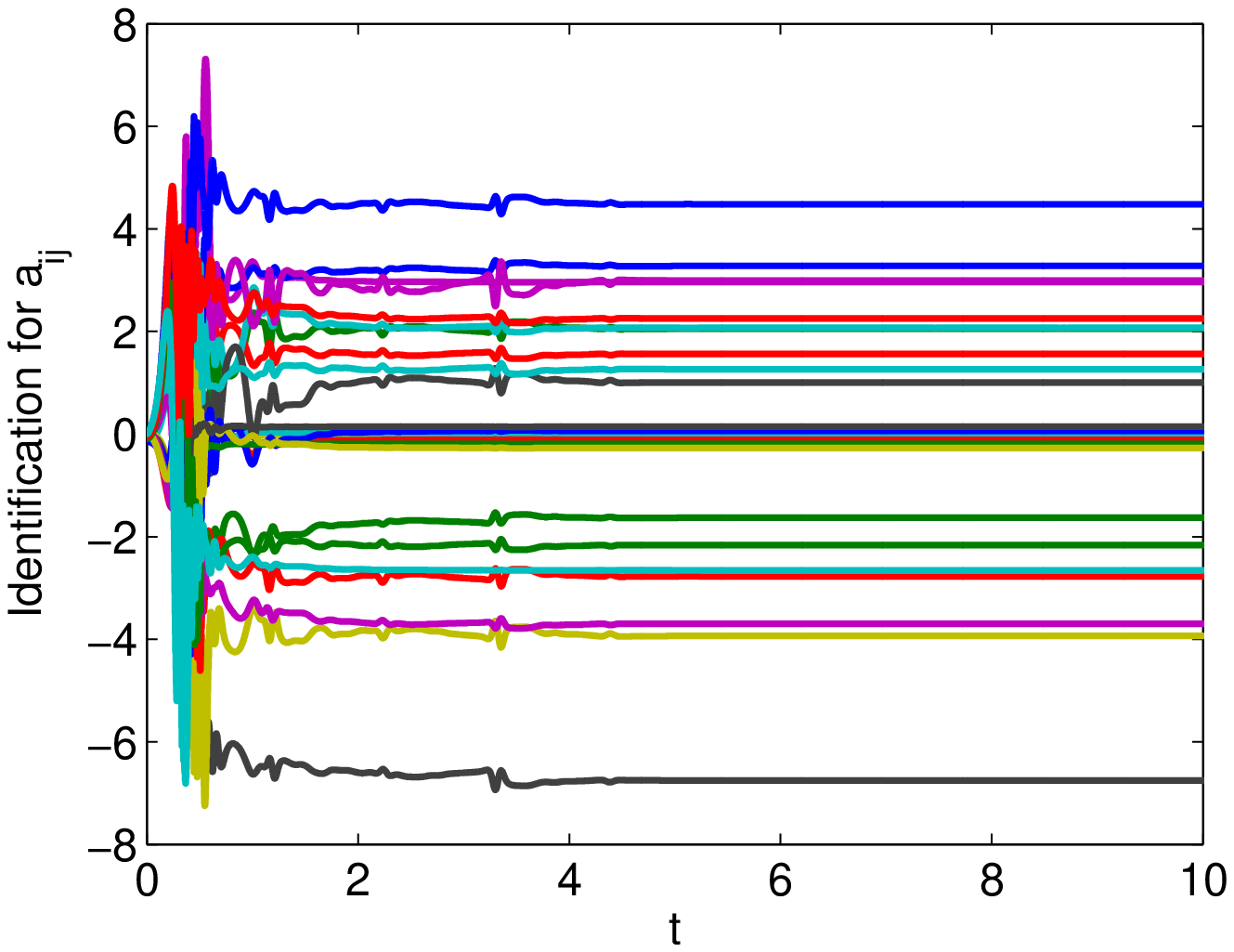}
\hspace*{1ex}
\includegraphics[height=5.0cm]{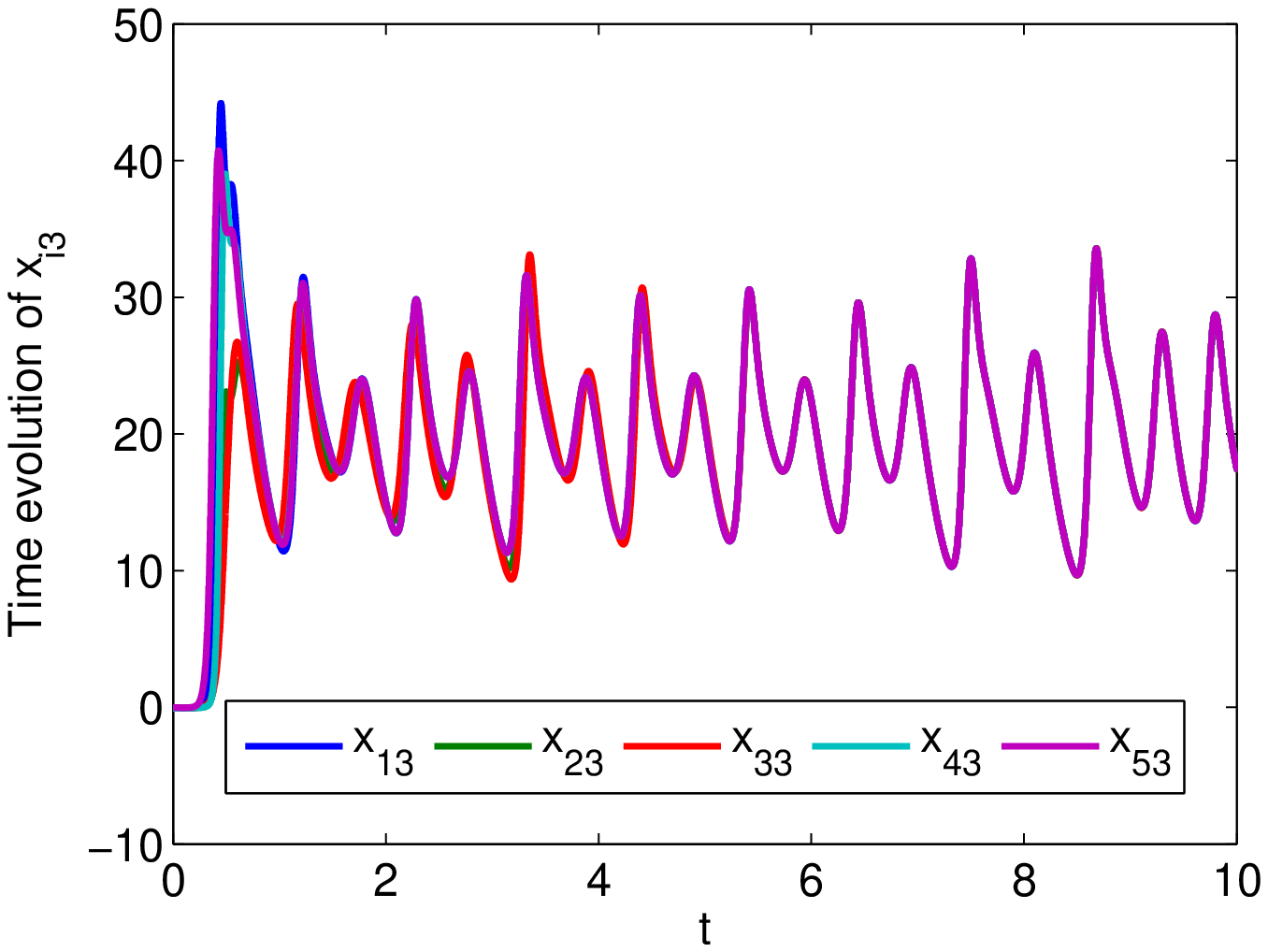}
\caption{Identification failure due to unsatisfied linear independence condition.  $\hat{a}_{ij}~(i,j=1,2,...,5)$ (left) and time evolution of the component variables $x_{i3}~(i=1,2,...,5)$ in the drive layer (right)  of the two-layer  L\"{u}-Chua network.}
\label{fig:ide_fail}
\end{figure*}

\begin{figure*}[!t]
\centering
\includegraphics[height=5.0cm]{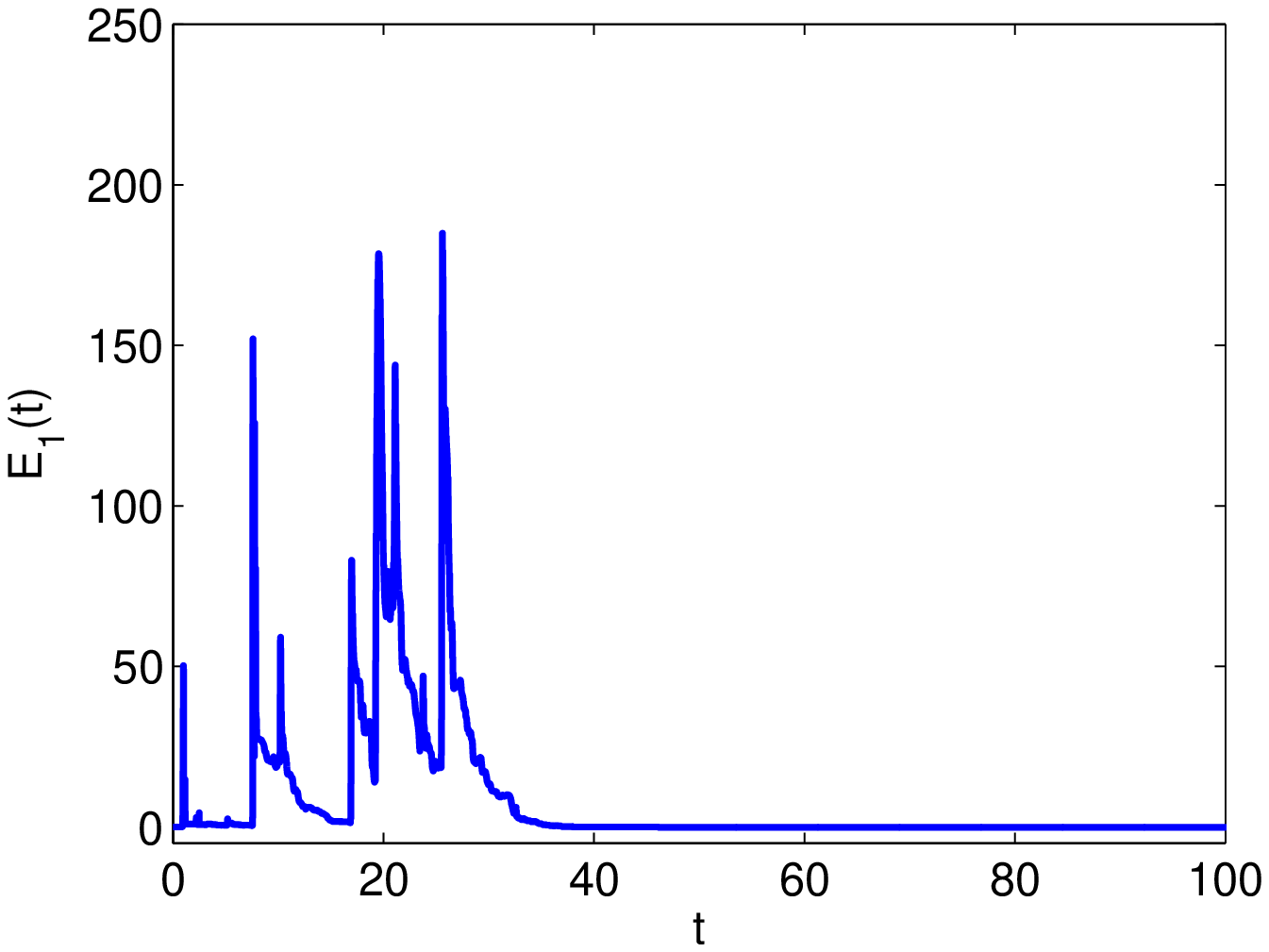}
\hspace*{1ex}
\includegraphics[height=5.0cm]{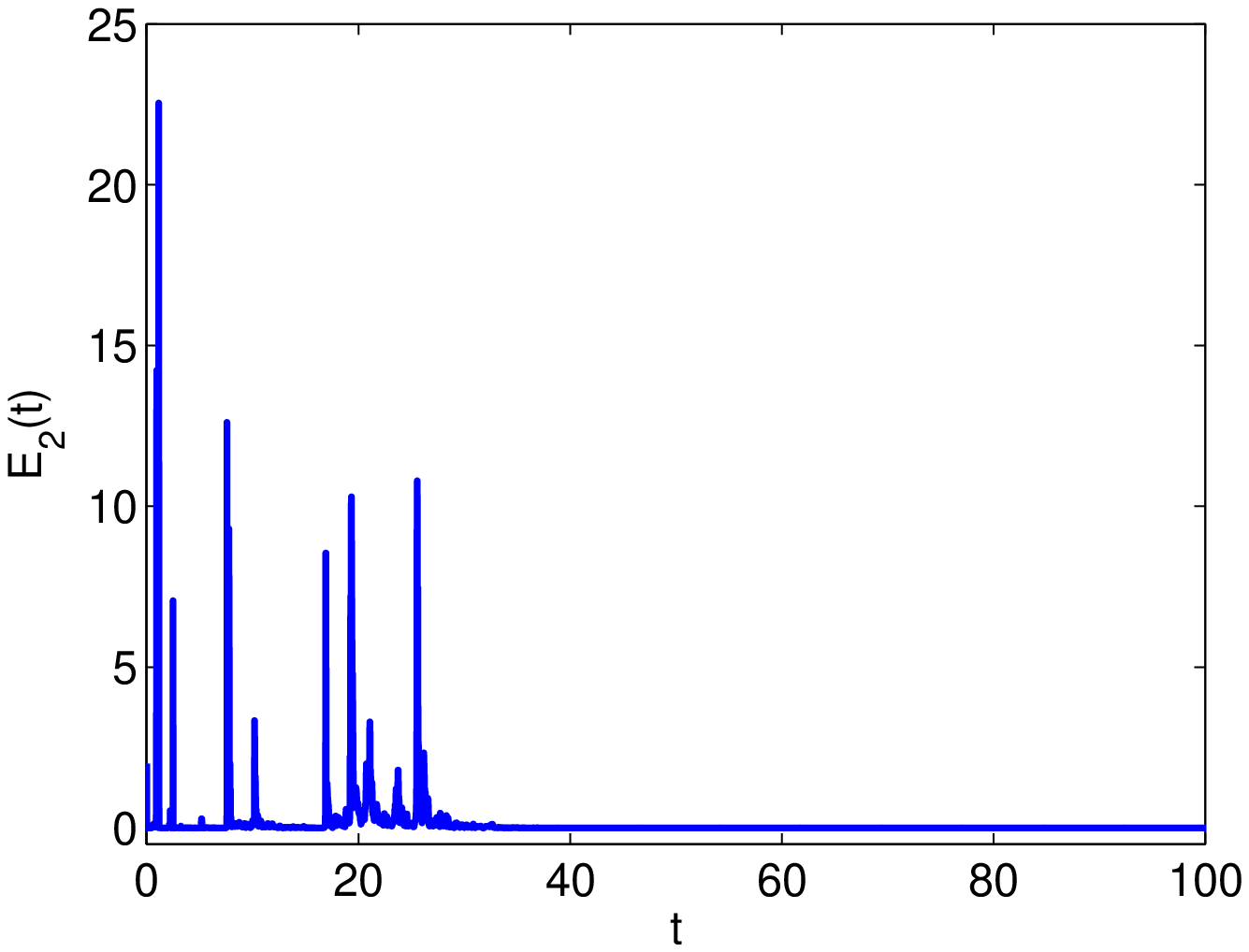}
\caption{Identification error (left) and synchronization error (right) of the two-layer hyperchaotic L\"{u}-Duffing network. }
\label{fig:err_4}

\vspace*{1.5ex}

\centering
\includegraphics[height=5.0cm]{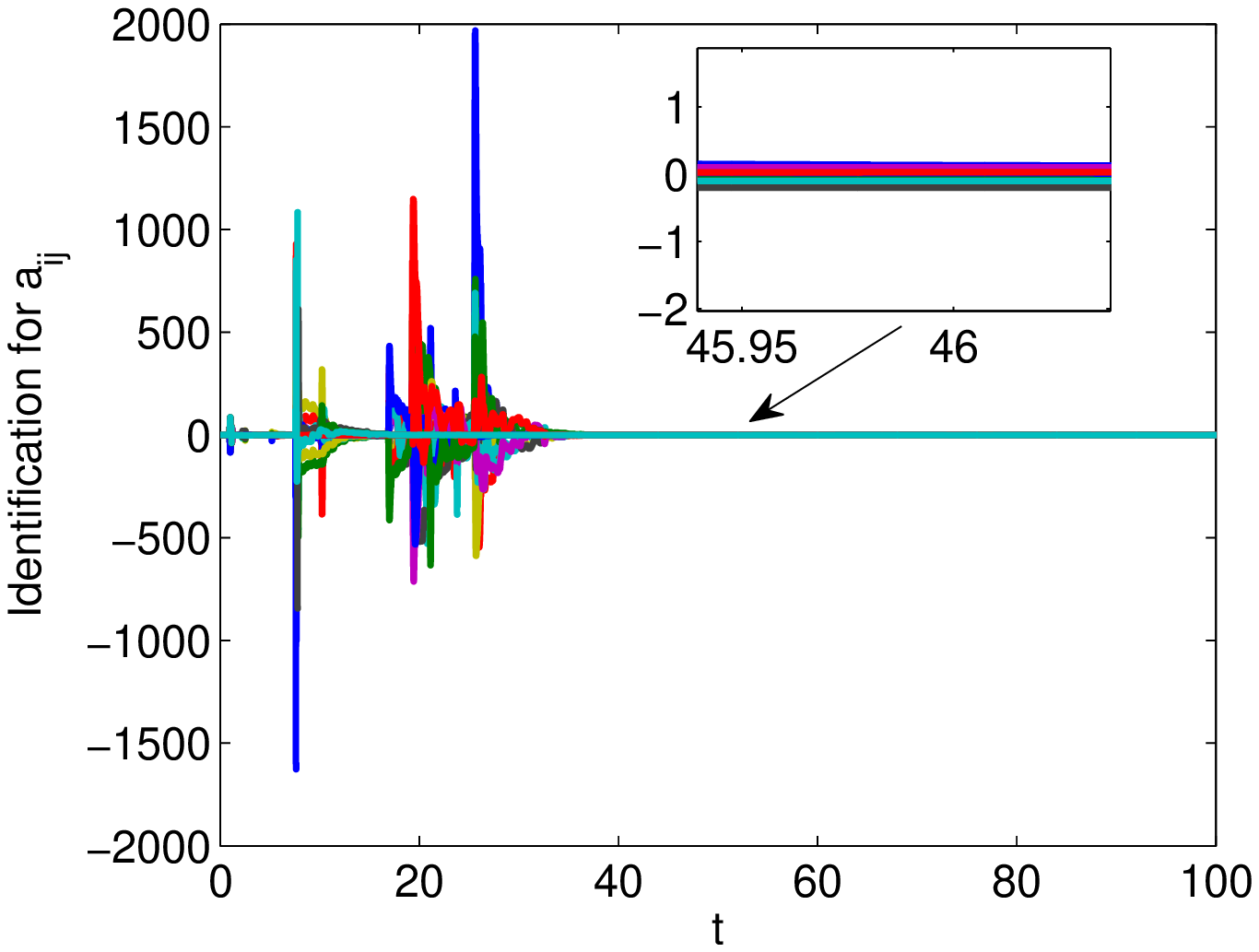}
\hspace*{1ex}
\includegraphics[height=5.0cm]{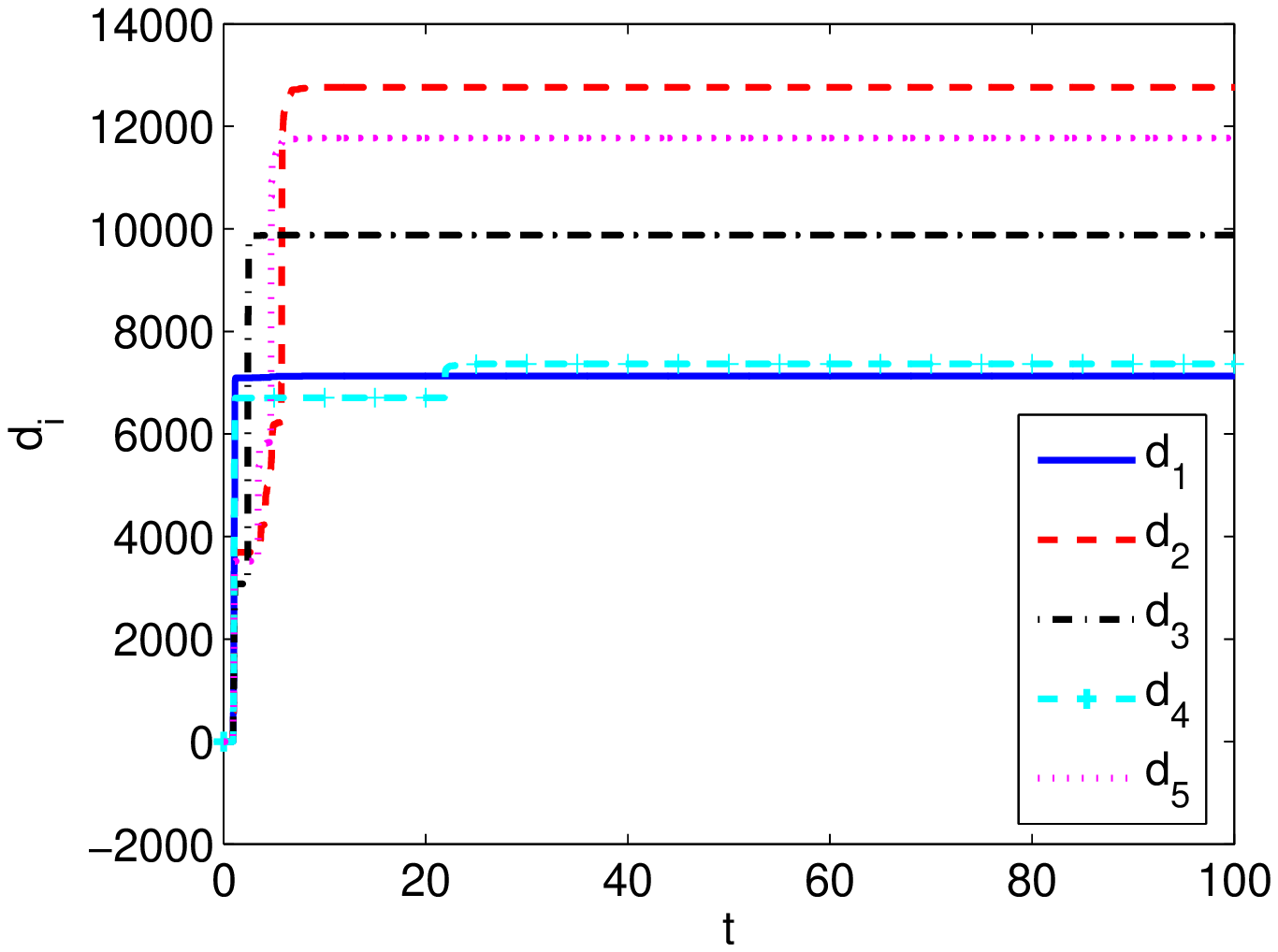}
\caption{Time evolution of $\hat{a}_{ij}~(i,j=1,2,...,5)$ (left) and    adaptive feedback gains $d_i~(i=1,2,...,5)$ (right) of   the two-layer hyperchaotic L\"{u}-Duffing network.}
\label{fig:aij_4}
\end{figure*}

\begin{figure}[htbp]
\centering
\includegraphics[width=8cm]{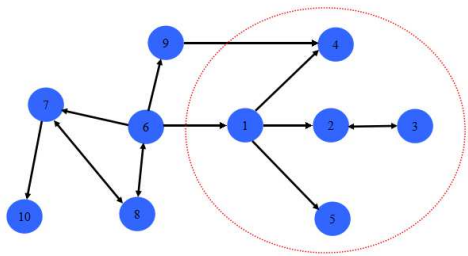}\\
{\caption{A testing network with 10 nodes, where the part in the red circle is the subnetwork of interest whose topology is   to be inferred. }}\label{fig:testingnetwork}
\end{figure}

\begin{figure*}[!t]
\centering
\includegraphics[height=5.0cm]{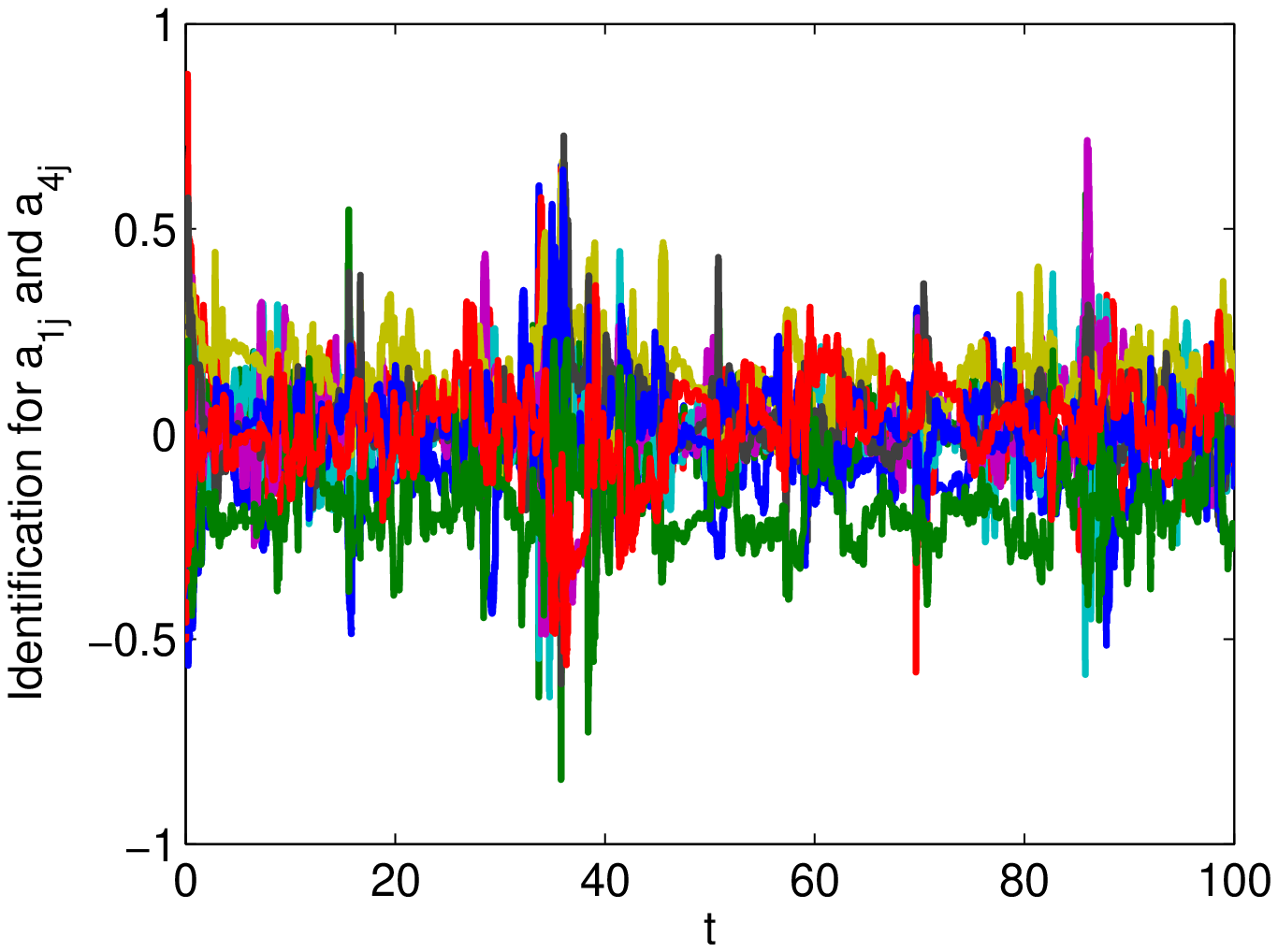}
\hspace*{1ex}
\includegraphics[height=5.0cm]{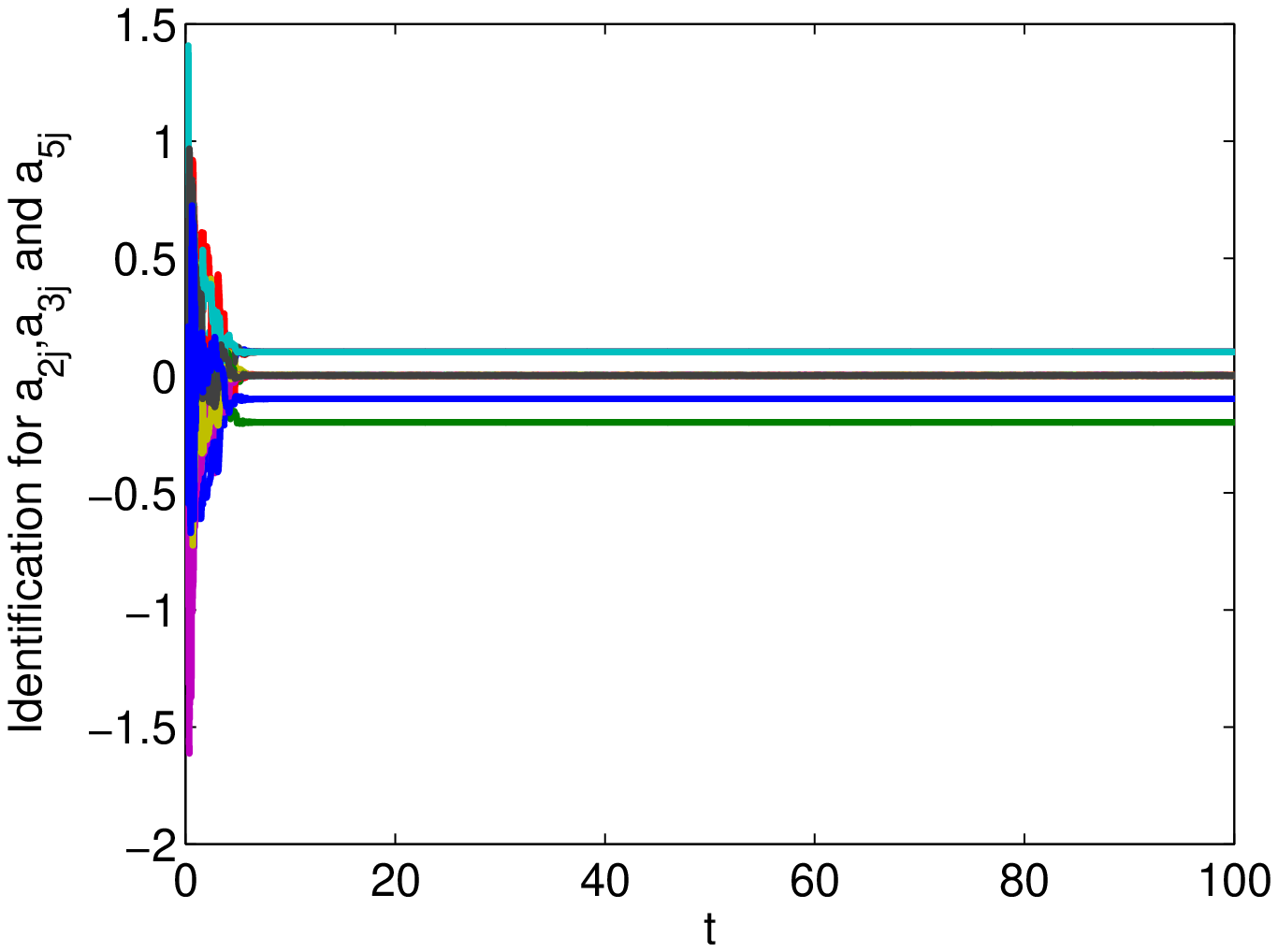}
\caption{Topology inference of a subnetwork. Left: time evolution of $\hat{a}_{ij} (i=1,4, j=1,2,\cdots, 5)$; right:  $\hat{a}_{ij} (i=2,3,5, j=1,2,\cdots, 5)$.}
\label{fig:aij_5}
\end{figure*}

In this section, numerical examples are presented to illustrate the effectiveness of the proposed identification schemes.
 Several benchmark chaotic systems, such as   L\"{u} system, Chua's circuit, Duffing system and hyperchaotic L\"{u} system,  will be taken as node dynamics. In fact, the well-known L\"{u} system and Chua circuit are respectively simplified models of several three dimensional physical systems, each of them having a chaotic attractor with appropriate system parameters. Moreover, L\"{u} system  is a special case of the unified chaotic system, which was proposed by L\"{u} \textit{et al.} in 2002 \cite{lu2002bridge}.  The unified chaotic system is described by
\begin{equation}\label{eq:unified chaotic}
\dot{\bx}_i=\left(\begin{array}{ccc}(25\theta+10)(x_{i2}-x_{i1})\\
(28-35\theta)x_{i1}-x_{i1}x_{i3}+(29\theta-1)x_{i2}\\
x_{i1}x_{i2}-\frac{\theta+8}{3}x_{i3}\end{array}
 \right),
\end{equation}
where $\theta \in [0,1]$. When $\theta=0.8$, system (\ref{eq:unified chaotic}) is  the L\"{u} system, and it reduces to the Lorenz system {\cite{lorenz1963deterministic}} when $\theta=0$ and  Chen system  {\cite{chen1999yet}} when $\theta=1$. Chua's circuit \cite{chua1992genesis} can be described as
\begin{equation}\label{eq:Chua's circuit}
\mathbf{\dot{x}}_i=\left(\begin{array}{ccc}\zeta(-x_{i1}+x_{i2}-l(x_{i1}))\\
 x_{i1}-x_{i2}+x_{i3}\\
 -\varrho x_{i2}\end{array}
 \right),
\end{equation}
where $l(x_{i1})=bx_{i1}+\frac{a-b}{2}(|x_{i1}+1|-|x_{i1}-1|)$ is a piecewise-linear function. It has a typical   double-scroll chaotic attractor for $\zeta=10, \varrho=18, a=-4/3$ and $b=-3/4$.

Duffing system is a typical two-dimensional chaotic system, as represented by
\begin{equation}\label{eq:duffing}
\dot{\bx}_i=\left(\begin{array}{ccc}x_{i2}\\
-p_2x_{i1}-p_3x_{i1}^3-p_1x_{i2}+q\cos \omega t\end{array}
 \right).
\end{equation}
It has a chaotic attractor when $p_1=0.4, p_2=-1.1, p_3=2, q=0.6$ and $\omega=1.8$.

There are many hyperchaotic systems emerging in the high dimensional social and economical systems. With high capacity, high security and high efficiency, hyperchaotic systems have been broadly used in secure communications, nonlinear circuits, biological systems and so on. Hyperchaotic L\"{u} system is a typical example of   high dimensional chaotic systems, which  is described by
\begin{equation}\label{eq:hyper lv}
\dot{\bx}_i=\left(\begin{array}{ccc}a(x_{i2}-x_{i1})+x_{i4}\\
-x_{i1}x_{i3}+cx_{i2}\\
x_{i1}x_{i2}-bx_{i3}\\
x_{i1}x_{i3}+dx_{i4}\end{array}
 \right).
\end{equation}
When $a=36, b=3, c=20$, system (\ref{eq:hyper lv}) has a hyperchaotic attractor for $-0.35<d\leq 1.30$, a chaotic attractor for $-0.46<d\leq -0.35$, and a periodic orbit for $-1.03<d\leq -0.46$.

In what follows, a classical five-node directed network is employed as the underlying testing network. The weighted coupling matrix is described as
\begin{equation*}
A=\rho
\left(
\begin{array}{ccccc}
0 & 0 & 0 & 0 & 0 \\
1 & -2 & 1 & 0 & 0 \\
0 & 1 & -1 & 0 & 0 \\
1 & 0 & 0 & -1 & 0 \\
1 & 0 & 0 & 0 & -1
\end{array}
\right),
\end{equation*}
where $\rho$ is the coupling strength, which is supposed to be 0.1 in the following simulations.

The stochastic differential delay equations are numerically solved employing the fourth-order Runge-Kutta method. The stochastic perturbations are randomly assigned and initial values of all the variables are set to be zeros. For brevity, the coupling delay is assumed to be 0.5, and the inner coupling matrix $H$ is supposed to be  an identity matrix with  a proper dimension.  The noise intensity function {is assumed to be $\p_i(t)=0.1 diag(e_{i1}(t),e_{i2}(t),\cdots,e_{in}(t)), i=1,2,\dots,N$, then one has  $\psi_i^T(t)\psi_i(t)=0.01 diag({e_{i1}^2(t),e_{i2}^2(t),\cdots,e_{in}^2(t)}).$ Therefore, one further obtains $\mbox{trace}(\psi_i^T(t)\psi_i(t))=0.01\Sigma_{j=1}^n e_{ij}^2(t)=0.01\mathbf{e}_i^T(t)\mathbf{e}_i(t)$, which  satisfies Assumption \ref{asp:noise bound}.}  {The identification error and synchronization error  are defined as} $E_1(t)=\sum\limits_{i=1}^N\sum\limits_{j=1}^N|\hat{a}_{ij}-a_{ij}|/N^2$ and {$E_2(t)=\sum\limits_{i=1}^N\|\mathbf{e}_i(t)\|_1/N$, respectively, where $\|\cdot\|_1$ represents 1-norm.}

$\mathbf{Example 4.1}$  Consider the L\"{u} system as  node dynamics in the drive network layer, and the Chua's circuit as node dynamics in the  response layer. It is obvious that Assumption \ref{asp:lipuxizi} is satisfied {\cite{li2006estimating}}. The map $\ph_i$ is taken as   $\by_i(t)=\ph_i(\bx_i(t))=(2x_{i1},x_{i2},\frac{1}{2}x_{i3}^2)$, then
\begin{equation*}
D\ph_i(\bx_i)=
\left(
\begin{array}{ccc}
2 & 0 & 0 \\
0 & 1 & 0 \\
0 & 0 & x_{i3}
\end{array}
\right).
\end{equation*}
The adaptive controllers and updating laws are designed according to (\ref{eq:updating laws3})-(\ref{eq:updating laws2}),  with parameters being  $k_i=30, \delta_{ij}=20$.

Figures \ref{fig:err_2}-\ref{fig:di_2} present successful identification results  upon {generalized  synchronization}. Identification error is displayed in the left panel of Fig. \ref{fig:err_2}, which illustrates that the underlying topology of the drive network layer (\ref{eq:drive}) is correctly inferred by the proposed control technique. The right panel of Fig. \ref{fig:err_2} presents time evolution of the generalized synchronization error, which goes to zero rapidly.  The left panel of Fig. \ref{fig:aij_2} gives time evolution of $\hat{a}_{ij}~(i,j=1,2,...,5)$. To have a clearer view, estimation of incoming edges of node 2 is given in the right panel. It is obvious that two curves corresponding to $\hat{a}_{21}$ and $\hat{a}_{23}$ get stabilized at 0.1,   two curves corresponding to $\hat{a}_{24}$ and $\hat{a}_{25}$ stabilize at  0,   and the curve $\hat{a}_{22}$  goes to $-0.2$. Figure \ref{fig:xi1xi3_2_2} shows the phase diagrams of node  2 in  the two-layer L\"{u}-Chua network. One can see that the two attractors are similar in a certain mode.  Furthermore,  the relationship between counterparts in the two network layers is examined. The left panel of Fig. \ref{fig:di_2} shows  $x_{i3}$ and $y_{i3}$ of node 2 in the two layers, where the transients are discarded.  The relationship is in accord with the predefined  generalized outer synchronization manifold $y_{i3}=\frac{1}{2}x_{i3}^2$.
The right panel of Fig. \ref{fig:di_2} displays  the time-varying adaptive feedback gains $d_i(t)~(i=1,2,...,5)$. It is seen that the adaptive feedback gains stabilize  at some constants upon successful topology identification. {Figure 5 displays the time evolution  of the component variables $x_{i3}~(i=1,2,...,5)$ of nodes in the drive layer for this successful identification case. For a clearer view, the time axis is restricted to the interval [0,10]. It is obvious that the trajectories are asynchronous.}

{It is well-known that the linear independence condition in Assumption \ref{asp:independent} is a very essential condition for guaranteeing successful topology identification. Just as mentioned in Remark \ref{rmk:linear independence condition},  synchronization in the unknown network will make the linearly independence condition unsatisfied, which further  leads to topology identification failure. In the case of identification failure, we usually start with checking whether nodes of the unknown network achieve some kind of synchronization, since this is more feasible and practical than checking the linear independence condition. This statement can be clearly illustrated by Figs. 5 and 6.  Figure 5 illustrates that  for successful topology identification, nodes are asynchronous.    Figure 6 displays the identification results for $a_{ij}~(i,j=1,2,...,5)$ and time evolution of the component variables $x_{i3}~(i=1,2,...,5)$ in the drive layer  of the two-layer  L\"{u}-Chua network for $\rho=5$ and $\tau=0.01$.  From the left panel, it is observed that although $\hat{a}_{ij}$ get stabilized,  they do not arrive at the  expected values. The reason can be explained from the right panel, where  the component variables $x_{i3}$  of each node in the unknown network  layer run into synchronization, which directly leads to the failure of topology identification. That is to say, synchronization hinders  topology identification.  }

$\mathbf{Example 4.2}$   Take the hyperchaotic L\"{u} system as  node dynamics in the drive network layer, and the Duffing system as   node dynamics in the response layer. The generalized synchronization map $\ph_i$ is supposed to be  $\by_i(t)=\ph_i(\bx_i(t))=(x_{i1}+2x_{i2},\frac{1}{2}x_{i3}+x_{i4})$, then
\begin{equation*}
D\ph_i(\bx_i)=
\left(
\begin{array}{cccc}
1 & 0 & 2 & 0 \\
0 & \frac{1}{2} & 0 & 1
\end{array}
\right).
\end{equation*}
The adaptive controllers and updating laws are designed accordingly.

Figure \ref{fig:err_4} displays  identification error  and synchronization error of  the two-layer hyperchaotic L\"{u}-Duffing network. It is obvious that the topological structures of the drive layer is successfully inferred and two layers reach predefined generalized outer synchronization.  Fig. \ref{fig:aij_4} shows   estimation for ${a}_{ij}~(i,j=1,2,...,5)$ and  time evolution of adaptive feedback gains $d_i~(i=1,2,...,5)$. It can be seen that  after transient oscillations,  $\hat{a}_{ij}$ get stabilized at $ {a}_{ij}$. Meanwhile, as can be obtained from the proof,  the adaptive feedback gains stabilize at  constants.

From this example, one can see that the proposed technique  can  be employed to infer unknown topologies of  complex networks composed of systems with high dimensions and complicated dynamics by constructing auxiliary networks composed of simpler systems.  From this viewpoint, {the proposed technique can greatly simplify practical design.}

\begin{figure*}[!t]
\centering
\includegraphics[height=5.0cm]{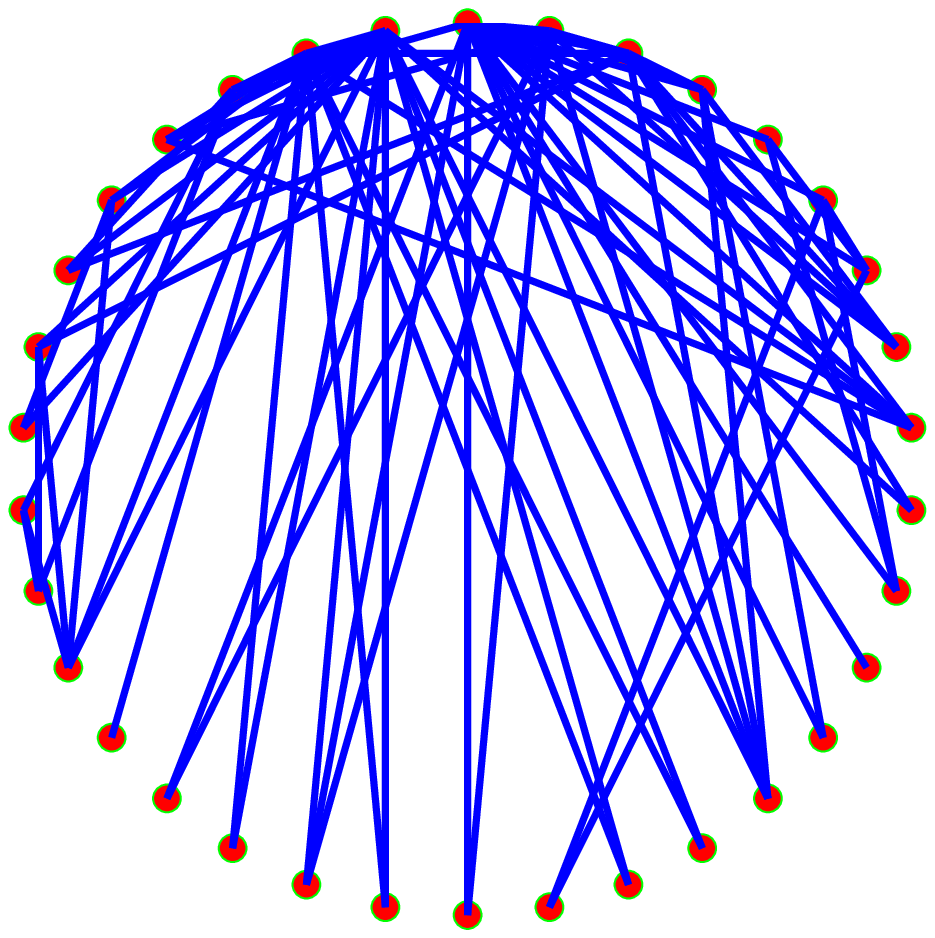}
\hspace*{1ex}
\includegraphics[height=5.0cm]{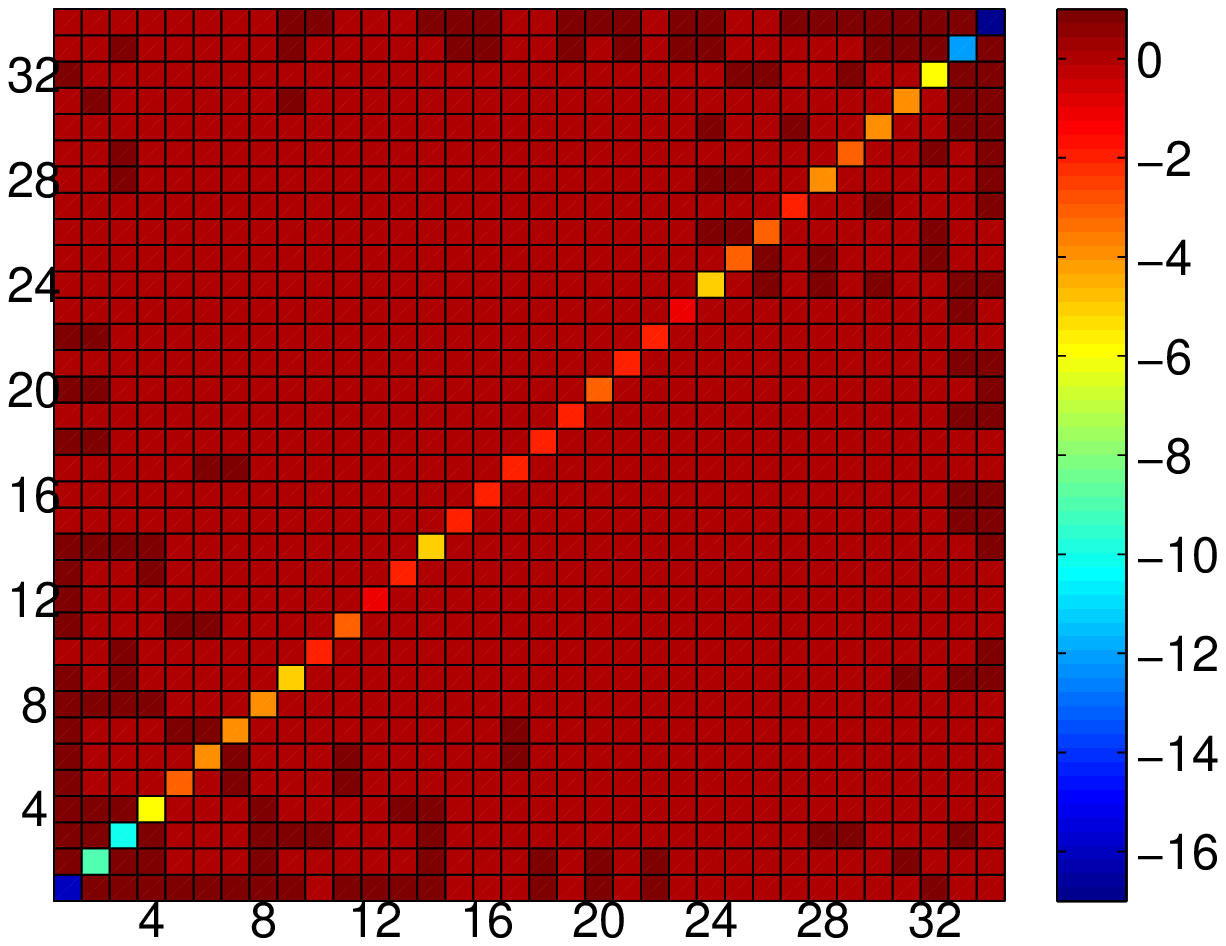}
{\caption{The model of the Zachary Karate Club network (left) and the colormap of the recovered unweighted coupling matrix (right).}}
\label{fig:testing_example4}

\vspace*{1.5ex}

\centering
\includegraphics[height=5.0cm]{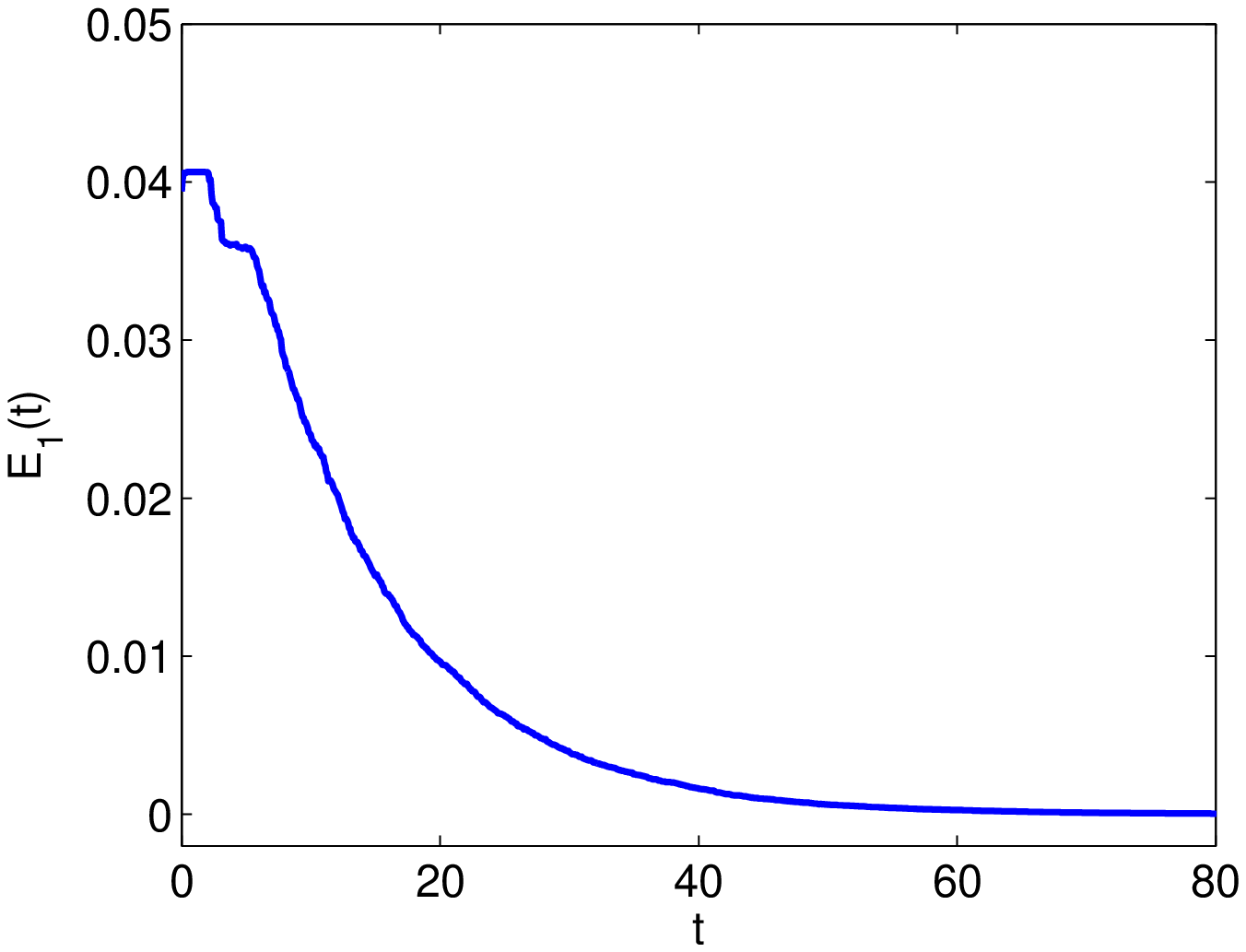}
\hspace*{1ex}
\includegraphics[height=5.0cm]{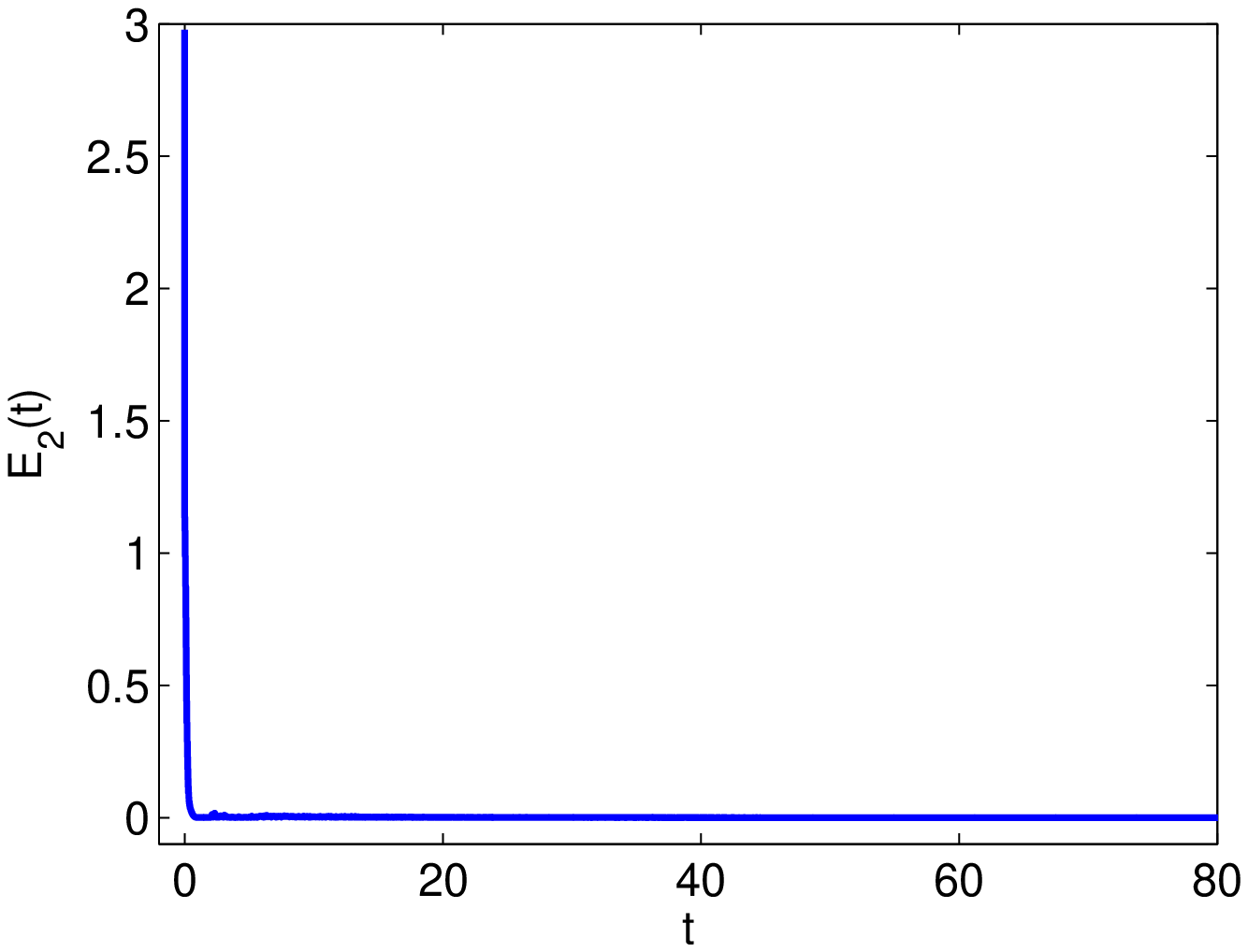}
{\caption{Identification error (left) and synchronization error (right) of the Zachary Karate Club network.}}
\label{fig:err_example4}
\end{figure*}

$\mathbf{Example 4.3}$ This method can also be employed to infer  connections in a subnetwork that is embedded within {a complex system.}  For simplicity, the network consisting of 10 unidirectionally connected nodes, as shown in Fig. 9,  is considered. Assume that our interest is to infer the topology of the subnetwork   composed of nodes 1 to 5 and each node's dynamical behavior can be monitored.  Nodes 1 and 4 are directly influenced by the hidden  nodes 6 and 9, respectively. To infer the  topology of this subnetwork, an auxiliary network consisting of five nodes is constructed. Let node dynamics of the interested 5-node subnetwork  be the L\"{u} system, and Chua's circuit as node dynamics of the auxiliary one. Controllers and updating laws are designed according to (\ref{eq:updating laws3})-(\ref{eq:updating laws2}), where   system parameters and  the map $\phi_i$ are supposed to be the same as those in Example 1.

Figure \ref{fig:aij_5} displays estimation for $ {a}_{ij}~(i,j=1,2,\cdots,5)$. It is shown that in  the left panel,  $\hat{a}_{1j}$ and $\hat{a}_{4j}$ are oscillating, while in the right panel, $\hat{a}_{2j}, \hat{a}_{3j}$ and $\hat{a}_{5j}$ get stabilized at  proper constants that are the exact corresponding values of $ {a}_{2j},  {a}_{3j}$ and $ {a}_{5j}$. This figure illustrates that the incoming links of nodes  which are free of latent disturbances can  be successfully inferred,  while nodes that they are directly  disturbed by hidden sources  cannot be identified and their estimation will oscillate accordingly. Therefore, the   method can be applied to infer topology of a subnetwork as well as locate the immediate neighbors of  hidden sources.

{$\mathbf{Example 4.4}$ Consider the well-known Zachary Karate Club network \cite{zachary1977information} as the testing network, which has 34 nodes and 78 edges, with the topology structure being displayed in the left panel of Fig. 11.  For brevity, take $\phi_i$ as a linear map in the numerical simulation, which is supposed to be  $\by_i(t)=\ph_i(\bx_i(t))=(2x_{i1}, x_{i2},  x_{i3}+1)$, then
\begin{equation*}
D\ph_i(\bx_i)=
\left(
\begin{array}{ccc}
2 & 0 & 0 \\
0 & 1 & 0 \\
0 & 0 & 1
\end{array}
\right).
\end{equation*}
The coupling strength $\rho$ of the unknown coupling matrix is supposed to be 0.01, and the coupling delay is  2. Adaptive controllers and updating laws are designed according to (\ref{eq:updating laws3})-(\ref{eq:updating laws2}), and the parameters $k_i$ in  (\ref{eq:updating laws2}) are taken as 100. Other parameters are designed as the same as those in Example 4.1. The right panel of Fig. 11 presents the the colormap of the recovered unweighted coupling matrix.   Figure 12 gives the identification error and synchronization error. It can be obtained that the unknown topology of the testing Zachary Karate Club network has been correctly recovered, and the drive layer achieves generalized synchronization with the constructed response layer. This further illustrates the effectiveness of the proposed method.
}

\section{Conclusions}\label{sec:con}

Topology identification of complex dynamical networks containing communication delay has been investigated via driving-based generalized synchronization of two-layer networks.  An adaptive control technique has been proposed to infer the underlying  topology of a dynamical network by constructing an auxiliary layer consisting of an identical number of systems.  The auxiliary layer is driven by signals from the unknown network so that it  reaches generalized outer synchronization with the drive layer and   successful topology identification   is achieved  in the sense of mean square. Since  perturbations caused by control input  have  been taken into consideration,  the proposed method is  comparatively more practical than previous results.
The main theorem contains some recent results as special cases.   Numerical simulations have been performed to illustrate the effectiveness of the proposed adaptive identification strategies.  Particularly, when the considered network is composed of systems with high-dimension or complicated dynamics, a much simpler auxiliary layer can be constructed for the purpose of topology inference.
Moreover, it has been shown that method can be applied to infer topology of a subnetwork embedded within {a network} and
locate hidden sources. Our results provide engineers {\color{blue}with certain theoretical supports and guidance for monitoring network structures as well as locating hidden sources.} We expect that our analysis could prompt attention and provide basic insight into further research endeavors on understanding practical and economical topology identification.

\section{Acknowledgments}
This work was supported by the National Natural Science Foundation of China under Grants 61573252 and 61203159, the Fundamental Research Funds for the Central Universities under Grant 2014201020206, and the Youth Fund Project of the Humanities and Social Science Research for the Ministry of Education of China under Grant 14YJCZH173.

\bibliographystyle{iop}
\bibliography{references}

\end{document}